\documentclass[aps,prd,twocolumn,groupedaddress,floatfix,showpacs,showkeys]{revtex4}

\usepackage[utf8]{inputenc}
\usepackage[T1]{fontenc}
\usepackage{slashed}
\usepackage{times}

\usepackage{amssymb,amsfonts,amsmath,amsthm}
\usepackage{dsfont,bbm}
\usepackage{dcolumn}

\usepackage{graphicx}
\usepackage[caption=false]{subfig}
\usepackage{color}

\newcommand{\up}{{\uparrow}}
\newcommand{\down}{{\downarrow}}

\newcommand{\tr}[1]{\operatorname{tr} \left\{ #1 \right\}}
\newcommand{\ide}[1]{\hskip #1 & \hskip -#1}

\begin{document}

\title{Light-cone Distribution Amplitudes of the Nucleon and Negative Parity Nucleon Resonances from Lattice QCD}

\author{V.M.~Braun}
   \affiliation{Institut f\"ur Theoretische Physik, Universit\"at Regensburg, 93040 Regensburg, Germany}
\author{S.~Collins}
   \affiliation{Institut f\"ur Theoretische Physik, Universit\"at Regensburg, 93040 Regensburg, Germany}
\author{B.~Gl\"a{\ss}le}
   \affiliation{Institut f\"ur Theoretische Physik, Universit\"at Regensburg, 93040 Regensburg, Germany}
\author{M.~{G\"ockeler}}
   \affiliation{Institut f\"ur Theoretische Physik, Universit\"at Regensburg, 93040 Regensburg, Germany}
\author{A.~{Sch\"afer}}
    \affiliation{Institut f\"ur Theoretische Physik, Universit\"at Regensburg, 93040 Regensburg, Germany}
\author{R.~W.~{Schiel}}
   \email{rainer.schiel@ur.de}
   \affiliation{Institut f\"ur Theoretische Physik, Universit\"at Regensburg, 93040 Regensburg, Germany}
\author{W.~{S\"oldner}}
    \affiliation{Institut f\"ur Theoretische Physik, Universit\"at Regensburg, 93040 Regensburg, Germany}
\author{A.~{Sternbeck}}
    \affiliation{Institut f\"ur Theoretische Physik, Universit\"at Regensburg, 93040 Regensburg, Germany}
\author{P.~{Wein}}
   \affiliation{Institut f\"ur Theoretische Physik, Universit\"at Regensburg, 93040 Regensburg, Germany}

\date{\today}

\begin{abstract}
We present the results of a lattice study of light-cone distribution amplitudes (DAs) of the nucleon and negative parity nucleon resonances 
using  two flavors of dynamical (clover) fermions on lattices of different volumes and pion masses down to $m_\pi\simeq 150$ MeV.
We find that the three valence quarks in the proton share their momentum in the proportion $37\%:31\%:31\%$, where the larger fraction corresponds 
to the $u$-quark that carries proton helicity, and determine the value of the wave function at the origin in position space, which turns out to
be small compared to the existing estimates based on QCD sum rules. Higher-order moments are constrained by our data and are all compatible with zero within our uncertainties.  
We also calculate the normalization constants of the higher-twist DAs 
that are related to the distribution of quark angular momentum. 
Furthermore, we use the variational method and customized parity projection operators 
to study the states with negative parity. 
In this way we are able to separate the contributions 
of the two lowest states that, as we argue, possibly correspond to $N^*(1535)$ and a mixture of $N^*(1650)$ and the pion-nucleon continuum, respectively. 
It turns out that the state that we identify with $N^*(1535)$ has a very different DA as compared to both the second observed negative parity state 
and the nucleon, which may explain the difference in the decay patterns  of $N^*(1535)$ and $N^*(1650)$ observed in experiment. 
\end{abstract}


\maketitle

\section{Introduction}

Understanding nucleon structure in terms of quarks and gluons is an important goal of Quantum Chromodynamics (QCD). 
The full nucleon wave function is very complicated and remains elusive but substantial progress was made for observables 
which require only specific limited nonperturbative input. 
In particular, hard exclusive reactions involving large momentum transfer from the initial to the final state baryon 
are dominated by the overlap of the light-cone wave functions at small transverse separations~\cite{Lepage:1980fj,Efremov:1979qk,Chernyak:1983ej} 
that are usually referred to as light-cone distribution amplitudes (DAs).  

The DAs are fundamental nonperturbative functions that are complementary to conventional parton distributions, but are much less well-known 
because their relation to experimental observables is less direct as compared to quark parton densities 
and on a more subtle theoretical footing. 
They are scale-dependent and for very large scales approach simple 
asymptotic expressions called asymptotic DAs~\cite{Lepage:1980fj,Chernyak:1983ej} that are widely believed, however, to provide one 
with a rather poor approximation for the momentum transfers accessible in modern experiments.

The theoretical description of DAs is based on the relation of their moments, i.e., integrals with powers of the momentum fractions, 
to matrix elements of local operators. 
Such matrix elements can be estimated using nonperturbative techniques, at least in principle, 
and the DAs reconstructed as an expansion in a suitable basis of polynomials in the momentum fractions. 
Historically, the first and the second moments of the nucleon DA have first been estimated using QCD 
sum rules~\cite{Chernyak:1984bm,King:1986wi,Chernyak:1987nu,Chernyak:1987nv} and the results indicated a very large deviation from the asymptotic
expressions. They were used extensively for model building of the DAs~\cite{Chernyak:1984bm,King:1986wi,Chernyak:1987nu,Chernyak:1987nv,Stefanis:1992nw} 
and allowed one to get a reasonable description of the experimental data on nucleon electromagnetic form factors and several other reactions 
within a purely perturbative framework, see, e.g., the review~\cite{Chernyak:1983ej}.

Despite a certain phenomenological success, this approach has remained controversial for many years. 
First, the QCD sum rules used to calculate the moments have been criticized as unreliable, see, e.g., \cite{Mikhailov:1986be}.
Second, it is commonly accepted nowadays that perturbative contributions to hard exclusive reactions at accessible energy scales 
must be complemented by the so-called soft or end-point corrections that correspond to a different (Feynman) mechanism to transfer the large 
momentum to a loosely bound system. Estimates of the soft contributions using QCD sum  rules, e.g., \cite{Radyushkin:1990te}, quark models~\cite{Bolz:1996sw}
and, more recently, light-cone sum rules~\cite{Braun:2001tj,Braun:2006hz,Anikin:2013aka} favor nucleon DAs that deviate from the asymptotic expressions
only mildly.  
 
With the advent of lattice QCD it has become possible to calculate moments of the DAs starting from first 
principles~\cite{Gavela:1988cp}, however, this task appears to be technically complicated so that detailed calculations 
are just beginning. 
The first quantitative results of lattice calculations of the moments of nucleon DAs have been 
obtained by the QCDSF collaboration~\cite{Gockeler:2008xv,Braun:2008ur} using  two flavors of dynamical (clover) fermions. 
The same group also made an exploratory study of the DAs of nucleon resonances with negative parity~\cite{Braun:2009jy}.
   
In this work we extend the analysis in \cite{Gockeler:2008xv,Braun:2008ur,Braun:2009jy} by making use of a much larger
set of lattices with different volumes, lattice spacings and pion masses down to $m_\pi\simeq 150$ MeV, 
and making various refinements in the procedure how the required matrix elements are extracted from lattice data. 
Our data allow one to perform, for the first time, a reliable chiral and infinite volume extrapolation of the results to the physical limit, 
and also a continuum extrapolation (to a lesser extent).

Our main results can be summarized as follows:      
\begin{itemize}
 \item We have calculated the nucleon coupling $f_N$ that corresponds to the probability amplitude to find the three valence quarks at one space point,
\begin{align}
 f_N &= 2.84(1)(33)\ 10^{-3}\ \text{GeV}^2\,.
\end{align}
Here and below all numbers refer to the scale $\mu^2 =4\ \text{GeV}^2$,
the first error is statistical, including the chiral extrapolation, and the second is due to the continuum extrapolation. This
number appears to be $\sim 30\%$ smaller than the existing estimates, which further decreases the
perturbative contribution to nucleon form factors. 
\item We have also calculated the nucleon couplings $\lambda_1$ and $\lambda_2$ that are related to the normalization of the $P$-wave three-quark wave functions
that involve orbital angular momentum 
\begin{align}
 \lambda_1 &= -4.13(2)(20)\ 10^{-2}\ \text{GeV}^2\,,
\notag\\
 \lambda_2 &= 8.19(5)(39)\ 10^{-2}\ \text{GeV}^2\,.
\end{align}
\item We have determined the momentum fractions carried by the three valence quarks in the proton
\begin{align}
 \langle x_1\rangle  &= 0.372(7)\,,
\notag\\
 \langle x_2\rangle  &= 0.314(3)\,,
\notag\\
 \langle x_3\rangle  &= 0.314(7)\,,
\label{result:<x>}
\end{align}
  where the first number corresponds to the $u$-quark that carries the proton helicity and the other two 
  to the $u,d$ quarks with helicities opposite to one another that are sometimes thought of as coupled in a scalar ``diquark''. 
  The approximate equality $\langle x_2\rangle \simeq \langle x_3\rangle $ was not expected
  and can be viewed as being consistent with the ``diquark'' picture.
\item We use the variational method and customized parity projection operators to study the states with negative parity. 
        In this way we are able to separate the contributions of the two lowest states that, as we argue, possibly correspond 
        to $N^*(1535)$ and a mixture of $N^*(1650)$ with the pion-nucleon continuum, respectively. 
        It turns out that the state that we identify with $N^*(1535)$ has a qualitatively 
        different DA compared to both the second observed negative parity state 
        and the nucleon: It has a very small value at the origin and is almost antisymmetric with respect to the interchange of the 
        quarks in the scalar ``diquark''. This result is important for the forthcoming studies of the electroproduction of nucleon resonances
        at large momentum transfers at the 12 GeV upgrade of the Jefferson Lab accelerator facility~\cite{Aznauryan:2012ba}
        and may explain the difference in the decay patterns  of $N^*(1535)$ and $N^*(1650)$ observed in experiment. 
\end{itemize}
 
The presentation is organized as follows. 
Section~\ref{sec:das} is introductory. We explain the relation between DAs and light-cone wave functions and introduce the required definitions and notations.
The necessary steps to compute moments of the DAs from lattice QCD are detailed in Section~\ref{sec:latdas}. 
The numerical analysis of our lattice data and their extrapolation to the physical point is presented in Section~\ref{sec:data}.
The final results are collected in Section~\ref{sec:res}, while Section~\ref{sec:concl} is reserved for the summary and conclusions.
The paper also contains several Appendices with a discussion of more technical issues.  

\section{Nucleon Wave Functions and Distribution Amplitudes} \label{sec:das}

The quantum-mechanical picture of a nucleon as a superposition of states with different numbers of partons is formulated in the infinite momentum frame or using light-cone quantization. 
Although \emph{a priori} there is no reason to expect that nucleon wave function components with, say, 100 partons (quarks and gluons) are suppressed as compared to those with only the three valence quarks, the phenomenological success of naive quark models suggests that only the first few  Fock components are relevant. At least in hard exclusive reactions which involve a large momentum transfer to the nucleon, the dominance of valence states is widely expected and can be proven within QCD perturbation theory~\cite{Lepage:1980fj,Chernyak:1983ej}.   

The most general parametrization of the three-quark sector involves six scalar light-cone wave functions ~\cite{Ji:2002xn,Ji:2003yj} which correspond to different possibilities to couple the quark helicities and the total orbital angular momentum to produce the helicity-$1/2$ nucleon state: $\lambda_1+\lambda_2+\lambda_3 + L_z = 1/2$. In particular, zero angular momentum is allowed, $L=0$, if the quark helicities $\lambda_i$ sum up to $1/2$. The corresponding contribution can be written as~\cite{Lepage:1980fj,Chernyak:1983ej,Ji:2002xn}:  
\begin{align}
\label{def:nucleonWF}
 |N(p)^\uparrow\rangle^{L=0} =& \frac{\epsilon^{abc}}{\sqrt{6}}
\int\frac{[dx][d^2\vec{k}]}{6\sqrt{x_1x_2x_3}}
\,\,\Psi_N(x_i,\vec{k}_i) |u_a^\uparrow(x_1,\vec{k}_1)\rangle
\nonumber\\
&{}\times \Big[
 \big|u_b^\downarrow(x_2,\vec{k}_2)\rangle |d_c^\uparrow(x_3,\vec{k}_3)\rangle
\nonumber\\
&{}\hspace*{0.5cm}
-\big|d_b^\downarrow(x_2,\vec{k}_2)\rangle |u_c^\uparrow(x_3,\vec{k}_3)\rangle
 \Big].
\end{align}
Here   
$\Psi_N(x_i,\vec{k}_i)$ is the  light-cone wave function that 
depends on the momentum fractions $x_i$ and the transverse momenta $\vec{k}_i$ of the quarks.
The integration measure is defined by 
\begin{align}
\int [dx] &= \int_0^1  dx_1 dx_2 dx_3\, \delta\big(\sum x_i-1\big)\,, \nonumber\\
\int [d^2\vec{k}] &= (16\pi^3)^{-2} \int  d^2\vec{k}_1 d^2\vec{k}_2 d^2\vec{k}_3\, \delta^2\big(\sum \vec{k}_i\big)\,.
\end{align}
In hard processes the contribution of $\Psi_N(x_i,\vec{k}_i)$ is dominant whereas the other existing three-quark wave functions give rise to a power-suppressed correction, i.e., a correction of higher twist.  

The light-front description of a nucleon is very attractive for model building, but faces conceptual difficulties that do not allow the calculation of light-cone wave functions from first principles, at least at present. 
In particular there are subtle issues related to renormalization and gauge dependence. 
An alternative approach describes nucleon structure in terms of distribution amplitudes corresponding to matrix elements of nonlocal gauge-invariant light-ray operators. 
The classification of DAs is based on twist rather than the number of constituents as for the Fock state wave functions. 
For example the leading-twist-three nucleon (proton) DA $\varphi_N(x_i)$ is defined by the matrix element~\cite{Braun:2000kw}  
\begin{align}
\label{varphi-N}
\ide{2em} \langle 0 | 
\epsilon^{ijk}\! \left(u^{\up}_i(a_1 n) C \!\!\not\!{n} u^{\down}_j(a_2 n)\right)  
\!\not\!{n} d^{\up}_k(a_3 n) 
|N(p)\rangle \nonumber\\
&= - \frac12 f_N\,p \cdot n\! \not\!{n}\, u_N^\up(p)\! \!\int\! [dx] 
\,e^{-i p \cdot n \sum x_i a_i}\, 
\varphi_N(x_i)\,,
\end{align}
where $q^{\up(\down)} = (1/2) (1 \pm \gamma_5) q$ are quark fields of given helicity, $p_\mu$ is the proton momentum, $p^2=m_N^2$, $u_N(p)$ the usual Dirac spinor in relativistic normalization, $n_\mu$ an auxiliary light-like vector $n^2=0$ and $C$ is the charge-con\-ju\-ga\-tion matrix. 
The relativistic normalization is tacitly assumed also for the state vector, $|N(p)\rangle$.
The Wilson lines that ensure gauge invariance are inserted between the quarks; they are not shown for brevity. The normalization constant $f_N$ is  defined in such a way that 
\begin{equation}
\label{norm}
  \int [dx]\, \varphi_N(x_i) =1\,.
\end{equation}
In principle, only the complete set of nucleon DAs carries the full information on the nucleon structure, in the same manner as the complete basis of light-cone wave functions. 
In practice, however, both expansions have to be truncated and the usefulness of a truncated version, taking into account either the first few Fock states or a few lowest twist 
contributions, may depend on the concrete physics application.
 
Using the wave function in Eq.~(\ref{def:nucleonWF}) to calculate the matrix element in Eq.~(\ref{varphi-N}) it is easy to show that the DA $\varphi_N(x_i)$ is related to the integral of the wave function $\Psi_N(x_i,\vec{k}_i)$ over transverse momenta, which corresponds to the limit of zero transverse separation between the quarks in position space~\cite{Lepage:1980fj}:
\begin{equation}
 f_N(\mu) \, \varphi_N(x_i,\mu) \sim \int\limits_{|\vec{k}|<\mu} [d^2\vec{k}]\, \Psi_N(x_i,\vec{k}_i)\,,
\end{equation}
where we have now explicitly stated the dependence on the scale $\mu$.
Thus, the normalization constant $f_N$ can be interpreted as the nucleon wave function at the origin (in position space).
 
As always in a field theory, extraction of the asymptotic behavior produces divergences that have to be regulated. 
As a result, the DAs become scheme- and scale-dependent. In the calculation of physical observables this dependence 
is cancelled by the corresponding dependence of the coefficient functions. 
The DA $\varphi_N(x_i,\mu)$ can be expanded in orthogonal polynomials $\mathcal{P}_{nk}(x_i)$ 
defined as eigenfunctions of the corresponding one-loop evolution equation:
\begin{equation}
   \varphi_N(x_i,\mu) = 120 x_1 x_2 x_3 \sum_{n=0}^\infty\sum_{k=0}^n \varphi^N_{nk}(\mu) \mathcal{P}_{nk}(x_i)\,,
\label{expand-varphi}
\end{equation}
where 
\begin{eqnarray}
  f_N(\mu) &=& f_N(\mu_0) \left(\frac{\alpha_s(\mu)}{\alpha_s(\mu_0)}\right)^{2/(3\beta_0)}\,,
\nonumber\\
  \varphi^N_{nk}(\mu) &=& \varphi^N_{nk}(\mu_0)\left(\frac{\alpha_s(\mu)}{\alpha_s(\mu_0)}\right)^{\gamma_{nk}/\beta_0}
\end{eqnarray}
and 
\begin{equation}
   \int [dx]\, x_1 x_2 x_3 \mathcal{P}_{nk}(x_i)\mathcal{P}_{n'k'}(x_i) \propto \delta_{nn'}\delta_{kk'}\,.
\end{equation}
Here $\beta_0 = 11-\frac23 n_f$ is the first coefficient of the QCD beta-function
and $\gamma_{nk}$ are the respective anomalous dimensions.  

The first few polynomials are
\begin{eqnarray}
\mathcal{P}_{00} &= & 1\,,
\nonumber\\
\mathcal{P}_{10} &= &21(x_1-x_3)\,,
\qquad
\mathcal{P}_{11} = 7(x_1-2x_2 + x_3)\,,
\nonumber\\
\mathcal{P}_{20} &= & \frac{63}{10}[3 (x_1-x_3)^2 -3x_2(x_1+x_3)+ 2x_2^2]\,,
\nonumber\\
\mathcal{P}_{21} & = &\frac{63}{2}(x_1-3x_2 + x_3)(x_1-x_3)\,,
\nonumber\\
\mathcal{P}_{22} & = & \frac{9}{5} [x_1^2\!+\!9 x_2(x_1\!+\!x_3)\!-\!12x_1x_3\!-\!6x_2^2\!+\!x_3^2]\,.
\label{lowest-P}
\end{eqnarray}
The corresponding anomalous dimensions are
\begin{align}
  \gamma_{00} =& 0\,, \qquad\quad\! \gamma_{10} = \frac{20}{9}\,,\qquad \gamma_{11} = \frac83\,,
\notag\\
 \gamma_{20} =& \frac{32}{9}\,,\qquad \gamma_{21} = \frac{40}{9}\,,\qquad \gamma_{22} =\frac{14}{3}\,.
\label{eq:adim3}
\end{align}
The normalization condition (\ref{norm}) implies that $\varphi^N_{00}=1$.
In what follows we will refer to the coefficients $\varphi^N_{nk}(\mu_0)$ as shape parameters. 
For a given order of the polynomials $n$, the coefficients $\varphi^N_{nk}\,, k =0,1,\ldots, n$ are 
ordered according to increasing anomalous dimension, cf.~Eq.~(\ref{eq:adim3}).   
They are related to the expansion coefficients used in Refs.~\cite{Braun:2008ur,Braun:2009jy} by
\begin{align}
\varphi^N_{10} = \frac1{21} c_{11}\,, && \varphi^N_{11} = \frac17 c_{10}\,,\notag\\
\varphi^N_{20} = \frac{10}{63} c_{22}\,, && \varphi^N_{21} = \frac{2}{63} c_{21}\,, && \varphi^N_{22} = \frac59 c_{20}\,.
\end{align}
The set of $\varphi^N_{nk}$ together with the normalization constant $f_N(\mu_0)$ at a certain reference scale $\mu_0$ 
specifies the momentum fraction distribution of the valence quarks in the nucleon. 
They are nonperturbative parameters that can be related to matrix elements of local gauge-invariant three-quark operators (see below).

In the last twenty years evidence has mounted that the simple-minded picture of a proton
with the three valence quarks in an S-wave is incomplete, so that for example the proton spin
is definitely not constructed from the quark spins alone and also the electromagnetic Pauli form factor
$F_2(Q^2)$ cannot be explained without quark orbital angular momentum contributions. 
The general classification of three-quark light-cone wave functions with nonvanishing angular momentum has been 
worked out in Refs.~\cite{Ji:2002xn,Ji:2003yj}. As shown in Ref.~\cite{Belitsky:2002kj}, the light-cone
wave functions with $L_z=\pm 1$ reduce, in the limit of small transverse separation,
to the twist-four nucleon DAs introduced in Ref.~\cite{Braun:2000kw}:
\begin{align} \label{twist-4}
\ide{2em} \langle 0 | \epsilon^{ijk}\! \left(u^{\up}_i(a_1 n) C\slashed{n} u^{\down}_j(a_2 n)\right) \!\slashed{p} d^{\up}_k(a_3 n) |N(p)\rangle \nonumber \\
=& -\frac14 \,p \cdot n\, \slashed{p}\, u_{N}^{\up}(p)\! \!\int\! [dx] \,e^{-i p \cdot n \sum x_i a_i}\, \nonumber \\
&\times \left[f_{N}\Phi^{N,WW}_4(x_i)+\lambda^N_1\Phi^{N}_4(x_i)\right], \nonumber\\
\ide{2em} \langle 0 | \epsilon^{ijk}\! \left(u^{\up}_i(a_1 n) C \slashed{n}\gamma_{\perp}\slashed{p} u^{\down}_j(a_2 n)\right) \gamma^{\perp}\slashed{n} d^{\up}_k(a_3 n) |N(p)\rangle \nonumber\\
=& -\frac12 \, p \cdot n\! \not\!{n}\,m_{N} u_{N}^{\up}(p)\! \!\int\! [dx] \,e^{-i p \cdot n \sum x_i a_i}\, \nonumber \\
&\times \left[f_{N}\Psi^{N,WW}_4(x_i)-\lambda^N_1\Psi^{N}_4(x_i)\right], \hspace*{2cm}\phantom{.} \nonumber\\
\ide{2em} \langle 0 | \epsilon^{ijk}\! \left(u^{\up}_i(a_1 n) C\slashed{p}\,\slashed{n} u^{\up}_j(a_2 n)\right) \!\not\!{n} d^{\up}_k(a_3 n) |N(p)\rangle \nonumber \\
=& \frac{\lambda^N_2}{12}\, p \cdot n\! \not\!{n}\, m_{N} u_{N}^{\up}(p)\! \!\int\! [dx] \,e^{-i p \cdot n \sum x_i a_i}\,\Xi^{N}_4(x_i) \,, 
\end{align}
where $\Phi^{N,WW}_4(x_i)$ and $\Psi^{N,WW}_4(x_i)$ are the so-called 
Wandzura-Wilczek contributions that can be expressed in terms of the 
leading-twist DA $\varphi_N(x_i)$~\cite{Braun:2008ia}:
\begin{align} \label{WW}
\Phi^{N,WW}_4(x_i) =& -\sum_{n,k}\frac{240\,\varphi^N_{nk}}{(n+2)(n+3)}
 \left(n+2-\frac{\partial}{\partial x_3}\right)
\nonumber\\ &{}\times x_1x_2x_3\mathcal{P}_{nk}(x_1,x_2,x_3)\,,
\nonumber\\
\Psi^{N,WW}_4(x_i) =& -\sum_{n,k}\frac{240\,\varphi^N_{nk}}{(n+2)(n+3)}
 \left(n+2-\frac{\partial}{\partial x_2}\right)
\nonumber\\ &{}\times x_1x_2x_3\mathcal{P}_{nk}(x_2,x_1,x_3)\,.
\end{align}
The two new constants $\lambda_1^N$ and $\lambda_2^N$
are defined in such a way that the integrals of the ``genuine'' twist-4 
DAs $\Phi_4$, $\Psi_4$, $\Xi_4$ are normalized to unity, similar to Eq.~(\ref{norm}).    
They have the same scale dependence to one-loop accuracy:
\begin{eqnarray}
  \lambda^N_{1,2}(\mu) &=& \lambda^N_{1,2}(\mu_0) \left(\frac{\alpha_s(\mu)}{\alpha_s(\mu_0)}\right)^{-2/\beta_0}.
\end{eqnarray}

The nonlocal operators entering the definitions of nucleon DAs 
do not have a definite parity. 
Thus the same operators couple also 
to the negative parity spin-1/2 nucleon resonances $N^*(1535)$, $N^*(1650)$, etc. 
One can define the leading-twist DA of these resonances from
\begin{align}
\ide{2em}
\langle 0 | \epsilon^{ijk}\! 
\left(u^{\up}_i(a_1 n) C \!\!\not\!{n} u^{\down}_j(a_2 n)\right)  
\!\not\!{n} d^{\up}_k(a_3 n) |N^*(p)\rangle \nonumber\\
&= \frac12 f_{N^*}\, p \cdot n\! \not\!{n}\, u_{N^*}^{\up}(p)\! 
\!\int\! [dx] \,e^{-i p \cdot n \sum x_i a_i}\, 
\varphi_{N^*}(x_i)\,,  \nonumber
\end{align}
where, of course, $p^2=m_{N^*}^2$. The constant $f_{N^*}$ has the physical 
meaning of the wave function of $N^*$ at the origin.  The 
DA $\varphi_{N^*}(x_i)$ is normalized to unity (\ref{norm}) and has the expansion
\begin{equation}
   \varphi_{N^*}(x_i,\mu) = 120 x_1 x_2 x_3 \sum_{n=0}^\infty\sum_{k=0}^n 
\varphi^{N^*}_{nk}(\mu) \mathcal{P}_{nk}(x_i)\,,
\label{expand-varphi-star}
\end{equation}
with the shape parameters $\varphi^{N^*}_{nk}$.

Similarly, there exist three independent subleading twist-4 
distribution amplitudes $\Phi^{N^*}_4$, $\Psi^{N^*}_4$,
$\Xi^{N^*}_4$ (as for the nucleon).
They can be defined as \cite{Braun:2009jy}
\begin{align}
\label{twist-4star}
\ide{2em}
\langle 0 | \epsilon^{ijk}\! 
\left(u^{\up}_i(a_1 n) C\slashed{n} u^{\down}_j(a_2 n)\right)  
\!\slashed{p} d^{\up}_k(a_3 n) |N^*(p)\rangle \nonumber\\
=& \frac14
\,p \cdot n\, \slashed{p}\, u_{N^*}^{\up}(p)\! \!\int\! [dx] 
\,e^{-i p \cdot n \sum x_i a_i}\,
\nonumber\\
&\times \left[f_{N^*}\Phi^{N^*,WW}_4(x_i)+\lambda^*_1\Phi^{N^*}_4(x_i)\right],   
\nonumber\\
\ide{2em}
\langle 0 | \epsilon^{ijk}\! 
\left(u^{\up}_i(a_1 n) C \slashed{n}\gamma_{\perp}\slashed{p} u^{\down}_j(a_2 n)\right)  
\gamma^{\perp}\slashed{n} d^{\up}_k(a_3 n) |N^*(p)\rangle \nonumber\\
=&
-\frac12
\, p \cdot n\! \not\!{n}\,m_{N^*} u_{N^*}^{\up}(p)\! \!\int\! [dx] 
\,e^{-i p \cdot n \sum x_i a_i}\,
\nonumber\\
&\times \left[f_{N^*}\Psi^{N^*,WW}_4(x_i)-\lambda^*_1\Psi^{N^*}_4(x_i)\right],
\hspace*{2cm}\phantom{.}
\nonumber\\
\ide{2em}
\langle 0 | \epsilon^{ijk}\! 
\left(u^{\up}_i(a_1 n) C\slashed{p}\slashed{n} u^{\up}_j(a_2 n)\right)  
\!\not\!{n} d^{\up}_k(a_3 n) |N^*(p)\rangle \nonumber\\
=& \frac{\lambda^*_2}{12}\, p \cdot n\! \not\!{n}\, m_{N^*} u_{N^*}^{\up}(p)\! \!\int\! [dx] 
\,e^{-i p \cdot n \sum x_i a_i}\,\Xi^{N^*}_4(x_i) \,, 
\end{align}
where $\Phi^{N^*,WW}_4(x_i)$ and $\Psi^{N^*,WW}_4(x_i)$ are given by the 
same expressions (\ref{WW}) in terms of the expansion of the 
leading-twist DA $\varphi_{N^*}(x_i)$ as for the nucleon.

The asymptotic distribution amplitudes (at very large scales) for the 
nucleon and the resonances are the same: 
\begin{align}
& \varphi^{\rm as}(x_i) = 120 x_1 x_2 x_3\,,\quad \Phi^{\rm as}_4(x_i) =  24
x_1 x_2\,,\quad
\nonumber\\
& \Phi^{WW,{\rm as}}_4(x_i)= 24 x_1 x_2 (1+\frac{2}{3}(1-5x_3))\,,
\nonumber\\
& \Psi^{WW,{\rm as}}_4(x_i)= 24 x_1 x_3(1+\frac{2}{3}(1-5x_2))\,,
\nonumber\\
& \Xi_4(x_i) = 24 x_2 x_3\,,\quad \Psi^{\rm as}_4(x_i) = 24 x_1 x_3\,.
\end{align}
For the sake of completeness we also give the definitions of the
normalization constants in terms of matrix elements of local 
three-quark operators. 
For the nucleon
\begin{align} \label{eq:nucnormconst}
&\langle 0 |\epsilon^{ijk}\! 
\left(u_i C\slashed{n} u_j\right)\!(0)  
\gamma_5\slashed{n} d_k(0) |N\!(p)\rangle\, = f_{N} p \cdot n  \, \slashed{n}\, u_N\!(p),
\nonumber\\
&\langle 0 | \epsilon^{ijk}\! 
\left(u_i C\gamma_{\mu} u_j\right)\!(0)  
\gamma_5\gamma^{\mu} d_k(0) |N\!(p)\rangle\, = \lambda^N_1 m_{N} u_{N}\!(p),
\nonumber\\
&\langle 0 | \epsilon^{ijk}\! 
\left(u_i C\sigma_{\mu\nu} u_j\right)\!(0)  
\gamma_5\sigma^{\mu \nu} d_k(0) |N\!(p)\rangle\, = \lambda^N_2 m_{N} u_{N}\!(p),
\end{align}
and similarly for $N^*$
\begin{align} \label{eq:nstarnormconst}
& \langle 0 |\epsilon^{ijk}\! 
\left(u_i C\slashed{n} u_j\right)\!(0)  
\gamma_5\slashed{n} d_k(0) |N^*\!(p)\rangle
\nonumber \\
& \hspace*{4cm}
 = f_{N^*} p \cdot n \gamma_5 \slashed{n}\, u_{N^*}\!(p),
\nonumber\\
& \langle 0 | \epsilon^{ijk}\! 
\left(u_i C\gamma_{\mu} u_j\right)\!(0)  
\gamma_5\gamma^{\mu} d_k(0) |N^*\!(p)\rangle
\nonumber \\
& \hspace*{4.0cm} = \lambda_1^{N^*} m_{N^*} \gamma_5 u_{N^*}\!(p),
\nonumber
\\
& \langle 0 | \epsilon^{ijk}\! 
\left(u_i C\sigma_{\mu\nu} u_j\right)\!(0)  
\gamma_5\sigma^{\mu \nu} d_k(0) |N^*\!(p)\rangle
\nonumber \\ 
& \hspace*{4.0cm} = \lambda_2^{N^*} m_{N^*} \gamma_5 u_{N^*}\!(p).\hspace*{1cm}{}
\end{align}
 
\section{Distribution amplitudes and lattice QCD} \label{sec:latdas}

On the lattice one can calculate moments of the DAs, e.g., 
\begin{equation}
 \Phi^{lmn} = \int [dx]\, x_1^l x_2^m x_3^n \,\varphi(x_i) \,,
\label{phi_lmn}
\end{equation}
which are related to matrix elements of local three-quark 
operators with covariant derivatives, as explained below.
The normalization is such that $\Phi^{000} = 1$. 
Starting from this Section spacetime is Euclidian and we use the Weyl representation for the 
$\gamma$--matrices; our conventions follow~\cite{Braun:2008ur}.

A traditional classification of leading-twist three-quark operators (in continuum theory) 
corresponds to a vector, axial and tensor Lorentz structure of the $u$-quark pair: 
\begin{align}
\mathcal{V}^{\rho\bar{l}\bar{m}\bar{n}}_\tau (0) =& \epsilon^{abc} \left[ i^l D^{\bar{l}} u(0) \right]^a_\alpha \left( C \gamma^\rho \right)_{\alpha \beta} \hspace*{2.2cm}{}
\nonumber \\
&{} \times \left[ i^m D^{\bar{m}} u(0) \right]^b_\beta \left[ i^n D^{\bar{n}} (\gamma_5 d(0)) \right]^c_\tau, 
\nonumber\\
\mathcal{A}^{\rho\bar{l}\bar{m}\bar{n}}_\tau (0) =&
 \epsilon^{abc} \left[ i^l D^{\bar{l}} u(0) \right]^a_\alpha \left( C \gamma^\rho \gamma_5 \right)_{\alpha \beta} 
\nonumber \\
&{} \times \left[ i^m D^{\bar{m}} u(0) \right]^b_\beta \left[ i^n D^{\bar{n}} d(0) \right]^c_\tau, 
\nonumber\\
\mathcal{T}^{\rho\bar{l}\bar{m}\bar{n}}_\tau (0) =&
 \epsilon^{abc} \left[ i^l D^{\bar{l}} u(0) \right]^a_\alpha \left( C (-i \sigma^{\xi \rho}) \right)_{\alpha \beta} 
\nonumber \\
&{} \times \left[ i^m D^{\bar{m}} u(0) \right]^b_\beta \left[ i^n D^{\bar{n}} (\gamma_\xi \gamma_5 d(0)) \right]^c_\tau,
\label{eq:VAT}
\end{align}
where we tacitly assume taking the leading-twist part, i.e., symmetrization and subtraction of traces.
The multi-index $\bar{l} \equiv \lambda_1 \cdots \lambda_l$, $D^{\bar{l}} \equiv D^{\lambda_1} \dots D^{\lambda_l}$ (and similarly for 
$\bar{m}$ and $\bar{n}$) denotes the Lorentz structure associated with the covariant derivatives
$D_\mu=\partial_\mu -igA_\mu$, whereas the indices $l,m,n$ (without bars)
stand for the total number of covariant derivatives acting on the first, second and third quark, respectively.

Matrix elements of these operators define a set of couplings $V^{lmn}$, $A^{lmn}$, $T^{lmn}$,
\begin{align}
 \langle 0 |\mathcal{V}^{\rho\bar{l}\bar{m}\bar{n}}_\tau |N(p)\rangle &= -f_N V^{lmn} p^\rho p^{\bar{l}}p^{\bar{m}}p^{\bar{n}} u_{N,\tau}(p),
\nonumber\\
 \langle 0 |\mathcal{A}^{\rho\bar{l}\bar{m}\bar{n}}_\tau |N(p)\rangle &= -f_N A^{lmn} p^\rho p^{\bar{l}}p^{\bar{m}}p^{\bar{n}} u_{N,\tau}(p),
\nonumber\\
 \langle 0 |\mathcal{T}^{\rho\bar{l}\bar{m}\bar{n}}_\tau |N(p)\rangle &= 2f_N T^{lmn} p^\rho p^{\bar{l}}p^{\bar{m}}p^{\bar{n}} u_{N,\tau}(p),
\end{align}  
which can be viewed as moments of auxiliary nucleon DAs $V(x_1,x_2,x_3)$, $A(x_1,x_2,x_3)$, $T(x_1,x_2,x_3)$.
These DAs are often used in practical calculations. 

Identity of the two $u$--quarks implies the symmetry relations
\begin{equation}
  V^{lmn} = V^{mln}\,,\qquad A^{lmn} = - A^{mln}\,, \qquad T^{lmn} = T^{mln}\,.    
\end{equation}
In addition, the requirement that the nucleon has isospin 1/2 allows one to express all $T$--moments
in terms of $V-A$:
\begin{equation}
 2 T^{lmn} = (V-A)^{lnm}+(V-A)^{mnl}\,.
\end{equation} 
The nucleon DA moments (\ref{phi_lmn}) are recovered as
\begin{equation} \label{eq:varphi}
 \Phi^{lmn} = (V-A)^{lmn}\,.
\end{equation}

Note that the operators defined in Eqs.~(\ref{eq:VAT}) and (\ref{eq:varphi}) do not have definite isospin themselves. We define
\begin{equation}
 \mathcal{F}^{\rho\bar{l}\bar{m}\bar{n}}_\tau = \frac13
\Big[\mathcal{V}^{\rho\bar{l}\bar{m}\bar{n}}_\tau - \mathcal{A}^{\rho\bar{l}\bar{m}\bar{n}}_\tau
- \mathcal{T}^{\rho\bar{l}\bar{n}\bar{m}}_\tau\Big]  
\label{K-operator}
\end{equation} 
which is an isospin-1/2 operator: It is annihilated by the isospin raising operator which is easy to verify using Fierz identities.

Knowing the matrix elements of $\mathcal{F}^{\rho\bar{l}\bar{m}\bar{n}}_\tau$ is sufficient. With
\begin{equation}
 \langle 0 |\mathcal{F}^{\rho\bar{l}\bar{m}\bar{n}}_\tau |N(p)\rangle = -f_N \phi^{lmn} p^\rho p^{\bar{l}}p^{\bar{m}}p^{\bar{n}} u_{N,\tau}(p)
\end{equation}
one gets
\begin{equation} \label{eq:kappalmndef}
    \phi^{lmn} = \frac13 \Big[ (V-A)^{lmn} + 2 T^{lnm}\Big]
\end{equation}
and
\begin{equation} \label{eq:kappalmnres}
  \Phi^{lmn} = 2\phi^{lmn}-\phi^{nml}\,.
\end{equation}
The shape parameters of the nucleon DA (\ref{expand-varphi}) can be obtained 
from the set of moments $\phi^{lmn}$ as follows:
\begin{align}
\varphi_{10} &= \frac{3}{2} \left( \phi^{100} - \phi^{001} \right),
\nonumber\\
\varphi_{11} &= \frac{1}{2} \left( \phi^{100} - 2 \phi^{010} + \phi^{001} \right), 
\nonumber\\
\varphi_{20} &=  3 \left( \phi^{200} \!\!+ \phi^{002} \!\!- 
\phi^{011} \!\!- \phi^{110} \right) + 2 \phi^{020} \!\!- 6 \phi^{101},
\nonumber\\[1mm]
\varphi_{21} &= 3 \left( \phi^{200}\! - \phi^{002}\right) + 9 \left( \phi^{011} - \phi^{110} \right), 
\nonumber\\[1mm]
\varphi_{22} &=  \phi^{200} \!\!-6 \phi^{020} \!\!+ \phi^{002} \!\!+ 
9 \left( \phi^{011} \!\!+ \phi^{110} \right) - 12 \phi^{101}.
\label{eq:cnk}
\end{align}
Momentum conservation ($x_1 + x_2 + x_3 = 1$) implies the following constraints:
\begin{equation}
\phi^{lmn} = \phi^{(l+1)mn} + \phi^{l(m+1)n} + \phi^{lm(n+1)}.
\label{eq:constraints}
\end{equation}
These relations can be used to rewrite (\ref{eq:cnk}) in equivalent alternative 
representations.
This possibility should, however, be used with caution, as the momentum conservation 
(in this form) is a consequence of the Leibniz rule for derivatives that is 
only fulfilled to $\mathcal{O}(a)$ accuracy in lattice simulations, cf. Subsection \ref{subsec:context}.

For the next-to-leading twist DAs we only consider the operators without derivatives
\begin{align} \label{eq:lmops}
\mathcal{L}_\tau (0) & =  \epsilon^{abc} u (0)^a_\alpha (C \gamma^\rho)_{\alpha \beta} u (0)^b_\beta \left[ \gamma_5 \gamma_\rho d (0) \right]^c_\tau, \nonumber \\
\mathcal{M}_\tau (0) & =  \epsilon^{abc} u (0)^a_\alpha (C \sigma^{\mu \nu})_{\alpha \beta} u (0)^b_\beta \left[ \gamma_5 \sigma_{\mu \nu} d (0) \right]^c_\tau,
\end{align}
which yield the next-to-leading twist normalization constants $\lambda_1$ and $\lambda_2$ defined in 
Eqs.~(\ref{eq:nucnormconst}) and (\ref{eq:nstarnormconst}).

\subsection{Lattice operators}

%
\begin{table*}
\renewcommand{\arraystretch}{1.2}
\caption{Irreducibly transforming multiplets.} \label{Table:IrredTrans}
\begin{ruledtabular}
\begin{tabular}{cccc}
 & dimension 9/2 & dimension 11/2 & dimension 13/2 \\
  & (0 derivatives)   & (1 derivative)   & (2 derivatives) \\
\hline \\[-11pt] $\tau^{\underline{4}}_1$ & $\mathcal{B}^{(0)}_{1,i}, \mathcal{B}^{(0)}_{2,i}, \mathcal{B}^{(0)}_{3,i}, \mathcal{B}^{(0)}_{4,i}, \mathcal{B}^{(0)}_{5,i}$ & & $\mathcal{B}^{(2)}_{1,i},\mathcal{B}^{(2)}_{2,i},\mathcal{B}^{(2)}_{3,i}$ \\
$\tau^{\underline{4}}_2$ & & & $\mathcal{B}^{(2)}_{4,i},\mathcal{B}^{(2)}_{5,i},\mathcal{B}^{(2)}_{6,i}$ \\
$\tau^{\underline{8}}$ & $\mathcal{B}^{(0)}_{6,i}$ & $\mathcal{B}^{(1)}_{1,i}$ & $\mathcal{B}^{(2)}_{7,i},\mathcal{B}^{(2)}_{8,i},\mathcal{B}^{(2)}_{9,i}$ \\
$\tau^{\underline{12}}_1$ & $\mathcal{B}^{(0)}_{7,i}, \mathcal{B}^{(0)}_{8,i}, \mathcal{B}^{(0)}_{9,i}$ & $\mathcal{B}^{(1)}_{2,i}, \mathcal{B}^{(1)}_{3,i}, \mathcal{B}^{(1)}_{4,i}$ & $\mathcal{B}^{(2)}_{10,i},\mathcal{B}^{(2)}_{11,i},\mathcal{B}^{(2)}_{12,i},\mathcal{B}^{(2)}_{13,i}$ \\
$\tau^{\underline{12}}_2$ & & $\mathcal{B}^{(1)}_{5,i}, \mathcal{B}^{(1)}_{6,i}, \mathcal{B}^{(1)}_{7,i}, \mathcal{B}^{(1)}_{8,i}$ & $\mathcal{B}^{(2)}_{14,i},\mathcal{B}^{(2)}_{15,i},\mathcal{B}^{(2)}_{16,i},\mathcal{B}^{(2)}_{17,i},\mathcal{B}^{(2)}_{18,i}$
\end{tabular}
\end{ruledtabular}
\renewcommand{\arraystretch}{1.0}
\end{table*}
%

Discretization of space and time reduces the Lorentz symmetry of the continuum theory to the discrete hypercubic symmetry of a four-dimensional
lattice. Thus, additional mixing between discretized versions of continuum operators
becomes allowed, and this mixing has to be reduced as much as possible by choosing a suitable operator basis.  
To this end, the three-quark operators that appear in the calculation of DAs have to be classified according to their transformation properties under the spinorial hypercubic group. 
The irreducibly transforming multiplets of three-quark operators have been found in Refs.~\cite{Kaltenbrunner:2008pb,Kaltenbrunner:2008zz} and their structure is shown schematically in Table \ref{Table:IrredTrans}. 
The left column contains the list of the five irreducible spinorial representations. 
Each entry in the table corresponds to a multiplet of baryon operators; e.g., 
$\mathcal{B}^{(0)}_{7,i}, \mathcal{B}^{(0)}_{8,i}, \mathcal{B}^{(0)}_{9,i}$ correspond to the three independent dodecuplets ($i=1,2,\ldots,12$) of three-quark operators without derivatives which transform according to the $\tau^{\underline{12}}_1$ representation. 
Explicit expressions for all operators with up to two derivatives are given in Refs.~\cite{Kaltenbrunner:2008pb,Kaltenbrunner:2008zz}. We refer to them as KGS operators in what follows.

For example
\begin{align}
&\mathcal{B}^{(0)}_{7,1} = u_1 u_3 d_3\,, \qquad
\mathcal{B}^{(0)}_{7,2} = \frac{1}{\sqrt{2}} \left( u_1 u_3 d_4 + u_1 u_4 d_3 \right)\,, \hspace*{0.5cm}
\nonumber \\
&\mathcal{B}^{(0)}_{7,3} = u_1 u_4 d_4\,,\qquad \mathcal{B}^{(0)}_{7,4} = u_2 u_3 d_3\,, 
\nonumber \\
&\mathcal{B}^{(0)}_{7,5} = \frac{1}{\sqrt{2}} \left( u_2 u_3 d_4 + u_2 u_4 d_3 \right)\,,
\qquad
\mathcal{B}^{(0)}_{7,6} = u_2 u_4 d_4\,, 
\nonumber \\
&\mathcal{B}^{(0)}_{7,7} = u_3 u_1 d_1\,, \qquad
\mathcal{B}^{(0)}_{7,8} = \frac{1}{\sqrt{2}} \left( u_3 u_1 d_2 + u_3 u_2 d_1 \right)\,, 
\nonumber \\
&\mathcal{B}^{(0)}_{7,9} = u_3 u_2 d_2\,, \qquad
\mathcal{B}^{(0)}_{7,10} = u_4 u_1 d_1\,, 
\nonumber \\
&\mathcal{B}^{(0)}_{7,11} = \frac{1}{\sqrt{2}} \left( u_4 u_1 d_2 + u_4 u_2 d_1 \right)\,,\quad
\mathcal{B}^{(0)}_{7,12} = u_4 u_2 d_2\,, 
\end{align}
where, e.g., $u_3$ stands for the third component of the $u$-quark bispinor (in the Weyl representation).
The operators $\mathcal{B}^{(0)}_{8, i}$ ($\mathcal{B}^{(0)}_{9, i}$) can be obtained from $\mathcal{B}^{(0)}_{7, i}$ 
by exchanging the spinor indices of quark one and two (one and three).

The KGS operators can be mapped to certain components of the three-quark operators in the $\mathcal{V},\mathcal{A},\mathcal{T}$
basis. In our example, $\mathcal{B}^{(0)}_{7, i}, \mathcal{B}^{(0)}_{8, i}$ and $\mathcal{B}^{(0)}_{9, i}$ correspond to the combinations $\mathcal{V}+\mathcal{A}, \mathcal{V}-\mathcal{A}$ and $\mathcal{T}$, respectively. A little algebra yields 
\begin{align}
\left( 
\begin{array}{c} -\mathcal{B}^{(0)}_{8,4} \\ \mathcal{B}^{(0)}_{8,3} \\  -\mathcal{B}^{(0)}_{8,10} \\  \mathcal{B}^{(0)}_{8,9} \end{array} 
\right)_\tau &= 
\frac{1}{4} \left( -\gamma_3 (\mathcal{V}^3_\tau - \mathcal{A}^3_\tau)  + \gamma_4 (\mathcal{V}^4_\tau - \mathcal{A}^4_\tau ) \right), 
\nonumber \\
\left( 
\begin{array}{c} -\mathcal{B}^{(0)}_{9,4} \\ \mathcal{B}^{(0)}_{9,3} \\  -\mathcal{B}^{(0)}_{9,10} \\  \mathcal{B}^{(0)}_{9,9} \end{array} 
\right)_\tau &= 
\frac{1}{4} \left( -\gamma_3 \mathcal{T}^3_\tau + \gamma_4 \mathcal{T}^4_\tau \right), 
\label{eq:theops}
\end{align}   
and similar representations can be worked out for all other cases.

The relations of this type reveal that the particular combinations of $\mathcal{V},\mathcal{A},\mathcal{T}$ 
operators that appear on the r.h.s.\ transform according to a particular irreducible representation of the spinorial hypercubic group (so that they 
are ``good'' lattice operators, in principle), but they do not have definite isospin yet. Isospin-1/2 operators can easily be constructed,
however, from suitable combinations of the KGS operators belonging to the same representation, and they can be expressed in terms of the 
$\mathcal{F}$ operators defined in Eq.~(\ref{K-operator}). For the above example, e.g., taking the difference between the two 
given operators one obtains
\begin{equation}
\mathcal{O}^{000}_{B,0} \equiv - \gamma_3 \mathcal{F}^3 + \gamma_4 \mathcal{F}^4.
\end{equation}
Another suitable combination is~\cite{Braun:2008ur}
\begin{equation}
\mathcal{O}^{000}_{C,0} \equiv  - \gamma_1 \mathcal{F}^1 - \gamma_2 \mathcal{F}^2 +  \gamma_3 \mathcal{F}^3 +  \gamma_4 \mathcal{F}^4. 
\end{equation}   
Both operators, $\mathcal{O}^{000}_{B,0}$ and $\mathcal{O}^{000}_{C,0}$, transform according to the $\tau^{\underline{12}}_1$ representation.

The lattice operators with one and two derivatives are constructed in a similar fashion.
In the following, curly braces indicate symmetrization over indices, e.g., 
$\mathcal{F}^{\{12\}} = \frac{1}{2!} \left( \mathcal{F}^{12} + \mathcal{F}^{21} \right)$.
We use in our calculations three operators with one derivative ($l+m+n=1$) from the $\tau^{\underline{12}}_2$ representation,
\begin{align}
 \mathcal{O}^{lmn}_{A,1} =& - 2 \gamma_1 \gamma_2 \mathcal{F}^{\{12\}} + \gamma_1 \gamma_3 \mathcal{F}^{\{13\}} 
                           + \gamma_1 \gamma_4 \mathcal{F}^{\{14\}}
\nonumber \\
&{} \quad  - \gamma_2 \gamma_3 \mathcal{F}^{\{23\}} - \gamma_2 \gamma_4 \mathcal{F}^{\{24\}}, 
\nonumber \\
\mathcal{O}^{lmn}_{B,1} =& 2 \gamma_3 \gamma_4 \mathcal{F}^{\{34\}} + \gamma_1 \gamma_3 \mathcal{F}^{\{13\}} 
                          - \gamma_1 \gamma_4 \mathcal{F}^{\{14\}} 
\nonumber \\
&{} \quad  + \gamma_2 \gamma_3 \mathcal{F}^{\{23\}} - \gamma_2 \gamma_4 \mathcal{F}^{\{24\}}, 
\nonumber\\ 
\mathcal{O}^{lmn}_{C,1} =& - \gamma_1 \gamma_3 \mathcal{F}^{\{13\}} + \gamma_1 \gamma_4 \mathcal{F}^{\{14\}} 
+ \gamma_2 \gamma_3 \mathcal{F}^{\{23\}} 
\nonumber \\
&{} \quad  - \gamma_2 \gamma_4 \mathcal{F}^{\{24\}}, 
\end{align} 
and the only existing isospin-1/2 operator with two derivatives ($l+m+n=2$) from the $\tau^{\underline{4}}_2$ representation,
\begin{align}
 \mathcal{O}^{lmn}_{2} =& - \gamma_1 \gamma_2 \gamma_3 \mathcal{F}^{\{123\}} + \gamma_1 \gamma_2 \gamma_4 \mathcal{F}^{\{124\}} 
\nonumber \\
&{} \quad  - \gamma_1 \gamma_3 \gamma_4 \mathcal{F}^{\{134\}} + \gamma_2 \gamma_3 \gamma_4 \mathcal{F}^{\{234\}}. 
\end{align}

It turns out that the twist-four operators $\mathcal{L}$ and $\mathcal{M}$ which were defined in Eq.~(\ref{eq:lmops}) are already good lattice operators and transform according to the $\tau^{\underline{4}}_1$ representation.

\subsection{Correlation functions}

On the lattice we measure correlation functions of these operators with a smeared nucleon source $\mathcal{N}_\tau$, which will be discussed in detail in Subsection \ref{subsec:isophy}.  

For $\mathcal{O}^{000}_{B,0}$ as an example, the contributions of the lowest positive and negative parity states to 
such a correlation function read
\begin{widetext}
\begin{eqnarray}
\lefteqn{\langle \mathcal{O}^{000}_{B,0} (t, \vec{p})_\tau \bar{\mathcal{N}}(0, \vec{p})_{\tau'} \rangle = }
\notag\\
& = &\sqrt{Z} f_N (i p_3 \gamma_3 + E_N \gamma_4) (E_N \gamma_4 - i \vec{p} \cdot \vec{\gamma} + m_N ) \frac{e^{-E_N t}}{2 E_N} 
 + \sqrt{Z_*} f_* (i p_3 \gamma_3 + E_* \gamma_4) (-E_* \gamma_4 + i \vec{p} \cdot \vec{\gamma} + m_* ) \frac{e^{-E_* t}}{2 E_*} 
\nonumber \\
&& - \sqrt{Z} f_N (i p_3 \gamma_3\! -\! E_N \gamma_4) (- E_N \gamma_4 - i \vec{p} \cdot \vec{\gamma} + m_N ) \frac{e^{-E_N (T- t)}}{2 E_N} 
 - \sqrt{Z_*} f_* (i p_3 \gamma_3 \!-\! E_* \gamma_4) (E_* \gamma_4 + i \vec{p} \cdot \vec{\gamma} + m_* ) \frac{e^{-E_* (T- t)}}{2 E_*}.
\nonumber\\
\label{eq:CF1}
\end{eqnarray} 
\end{widetext}
%
Here $\vec{p}=\{p_1,p_2,p_3\}$ is the momentum and we use the shorthand notations $f_\ast = f_{N^\ast}$, $m_\ast = m_{N^\ast}$ etc., for the quantities related to the negative parity state, $N^\ast$. $\sqrt{Z}$ is a -- usually momentum-dependent -- factor that indicates the overlap of the smeared nucleon source with the ``physical'' nucleon on the lattice. We explain how to eliminate this unknown factor at the end of this Subsection.

In this work we are specifically interested in a clean separation of states with different parity. 
Note that the correlation function in (\ref{eq:CF1}) is a matrix with respect to the spinor indices $\tau$ and $\tau'$. 
For convenience we multiply this expression by $\gamma_4$ and try to find a parity projection operator in the form  $\gamma_{\pm} \equiv \frac{1}{2} (1 + k_\pm \gamma_4)$, cf.~\cite{Lee:1998cx}, with $k_\pm$ to be determined from the condition that positive and negative parity states are distinguished by propagating forwards and backwards in time. For definiteness let us  consider the forward movers. We get
\begin{align}
\ide{2em} \hspace*{-0.5cm}\langle \left( \gamma_4 \mathcal{O}^{000}_{B,0} (t, \vec{p}) \right)_\tau \bar{\mathcal{N}}(0, \vec{p})_{\tau'} 
(1 + k_\pm \gamma_4)_{\tau' \tau} \rangle \nonumber \\
=& \sqrt{Z} f_N (k_\pm p_3^2 + k_\pm E_N^2 + m_N E_N) \frac{e^{-E_N t}}{2 E_N} 
\nonumber \\
&{} + \sqrt{Z_*} f_* (- k_\pm p_3^2 - k_\pm E_*^2 + m_* E_* ) \frac{e^{-E_* t}}{2 E_*}\,.
\end{align} 
One sees that if $p_3 = 0$ (but $p_1$ and $p_2$ arbitrary) choosing $k_+ = m_* / E_*$ annihilates the negative parity contribution and thus extracts the positive parity (nucleon) state, and, vice versa, $k_- = - m_N / E_N$ projects onto the negative parity state. 
It turns out that, under certain restrictions for the momenta $\vec{p}$, the same choice yields the correct parity projection for all correlation functions we are interested in. 
In the following expressions we show the positive parity contributions only and abbreviate $k \equiv k_+$ and $E \equiv E_N (\vec{p})$:   
\begin{align}
C^{000}_{B,0} =& 
\langle (\gamma_4 \mathcal{O}^{000}_{B,0}(t, \vec{p}))_\tau (\overline{\mathcal{N}}(0, \vec{p}))_{\tau'} (\gamma_+)_{\tau' \tau} \rangle 
\nonumber\\
=& f_N \sqrt{Z_N} \frac{E(m_N + k E) + k p_3^2}{E} e^{-E t},
\nonumber\\
C^{000}_{C,0} =& 
\langle (\gamma_4 \mathcal{O}^{000}_{C,0}(t, \vec{p}))_\tau (\overline{\mathcal{N}}(0, \vec{p}))_{\tau'} (\gamma_+)_{\tau' \tau} \rangle 
\nonumber\\
=&  f_N \sqrt{Z_N} \frac{E(m_N + k E) + k (p_1^2 + p_2^2 - p_3^2)}{E} e^{-E t},
 \end{align}
 \begin{align}
C^{lmn}_{A,1} =& 
\langle (\gamma_4 \gamma_1 \mathcal{O}^{lmn}_{A,1}(t, \vec{p}))_\tau (\overline{\mathcal{N}}(0, \vec{p}))_{\tau'} (\gamma_+)_{\tau' \tau} \rangle 
\nonumber\\
=& \!-\! f_N \phi^{lmn}\! \sqrt{Z_N} p_1 \frac{\!E(m_N \!+\! k E) \!+\! k (2 p_2^2 \!-\! p_3^2)}{E} e^{-E t}\!\!,
\nonumber\\
C^{lmn}_{B,1} =& 
\langle (\gamma_4 \gamma_1 \mathcal{O}^{lmn}_{B,1}(t, \vec{p}))_\tau (\overline{\mathcal{N}}(0, \vec{p}))_{\tau'} (\gamma_+)_{\tau' \tau} \rangle 
\nonumber\\
=& f_N \phi^{lmn} \sqrt{Z_N} p_1 \frac{E(m_N + k E) + k p_3^2}{E} e^{-E t},
\nonumber\\
C^{lmn}_{C,1} =&
\langle (\gamma_4 \gamma_1 \mathcal{O}^{lmn}_{C,1}(t, \vec{p}))_\tau (\overline{\mathcal{N}}(0, \vec{p}))_{\tau'} (\gamma_+)_{\tau' \tau} \rangle 
\nonumber\\
=& - f_N \phi^{lmn} \sqrt{Z_N} p_1 \frac{E(m_N + k E) + k p_3^2}{E} e^{-E t},
 \end{align}
and
 \begin{align}
C^{lmn}_{2} =& 
\langle (\gamma_2 \gamma_3 \gamma_4 \mathcal{O}^{lmn}_{2}(t, \vec{p}))_\tau (\overline{\mathcal{N}}(0, \vec{p}))_{\tau'} (\gamma_+)_{\tau' \tau} \rangle 
\nonumber\\
=& - f_N \phi^{lmn} \sqrt{Z_N} p_2 p_3 \frac{E(m_N + k E) + k p_1^2}{E} e^{-E t}.
\nonumber\\
\end{align}
For example, in order to have a nonzero overlap with the ground states we must keep $p_2$ and $p_3$ nonvanishing but set $p_1=0$ for the case of $C^{lmn}_{2}$. 
With this choice the contribution from negative parity states with mass-to-energy ratio $m_* / E_*$ is completely eliminated.  
Contributions from excited negative parity states, which have a different ratio $m/E$, are not completely eliminated but strongly suppressed. 
Together with the suppression due to smearing and the suppression due to the exponential decay with a higher mass, the positive parity states will dominate the signal, as desired.

For the twist-four correlation functions, we find:
\begin{align}
C_L &=  \langle (\mathcal{L}(t, \vec{p}))_\tau (\overline{\mathcal{N}}(0, \vec{p}))_{\tau'} (\gamma_+)_{\tau' \tau} \rangle \nonumber \\
    &=  \lambda_1 m_N \sqrt{Z_N} \frac{m_N + k E}{E} e^{-E t}, \nonumber \\
C_M &=  \langle (\mathcal{M}(t, \vec{p}))_\tau (\overline{\mathcal{N}}(0, \vec{p}))_{\tau'} (\gamma_+)_{\tau' \tau} \rangle \nonumber \\
    &=  \lambda_2 m_N \sqrt{Z_N} \frac{m_N + k E}{E} e^{-E t}. 
\end{align}

In order to determine the coupling constants, we have to eliminate the $\sqrt{Z_N}$ from the above equations. We do this by considering yet another correlation function, that of the smeared nucleon interpolator with itself:
\begin{align}
C_N &=  \langle (\mathcal{N}(t, \vec{p}))_\tau (\overline{\mathcal{N}}(0, \vec{p}))_{\tau'} (\gamma_+)_{\tau' \tau} \rangle \nonumber \\
    &=  Z_N \frac{m_N + k E}{E} e^{-E t}.
\end{align}
Taking the following ratio will then yield the desired result:
\begin{equation}
\left. \frac{C^{000}_{B,0}}{\sqrt{2 C_N}}\right|_{\vec{p}=0} = \left. \frac{C^{000}_{C,0}}{\sqrt{2 C_N}}\right|_{\vec{p}=0} = f_N m_N e^{-m_N t / 2},
\end{equation}
and similarly for $\lambda_1$ and $\lambda_2$. Finally, the moments $\phi^{lmn}$ are best determined by taking the following ratios:
\begin{align}
\left. \frac{C^{lmn}_{A,1}}{C^{000}_{B,0}}\right|_{p_2 = p_3 = 0} = & - \left. \frac{C^{lmn}_{B,1}}{C^{000}_{B,0}}\right|_{p_2 = p_3 = 0} =  \left. \frac{C^{lmn}_{C,1}}{C^{000}_{B,0}}\right|_{p_2 = p_3 = 0} \nonumber \\
= & - \phi^{lmn} p_1, \nonumber \\
\left. \frac{C^{lmn}_2}{C^{000}_{C,0}}\right|_{p_1 = 0, p_2^2 = p_3^2} = & - \phi^{lmn} p_2 p_3. 
\end{align}

\subsection{Renormalization}

The set of operators belonging to a given representation is closed under renormalization. For $f_N$, we have
\begin{align}
f_N^r = Z^{f_N} f_N^{\text{lat}},
\end{align}  
where the renormalization constant $Z^{f_N}$ should not be confused with the $\sqrt{Z}$-factor from the previous Subsection, $r$ denotes the renormalized value and ``$\text{lat}$'' the lattice value. For $\lambda_{1,2}$, 
\begin{align}
\lambda_i^r = Z^\lambda_{ij} \lambda_j^{\text{lat}},
\end{align}
where a sum over repeated indices is implied, and for the moments of the distribution amplitude, $\phi^{(1)}_i = (\phi^{100}, \phi^{010}, \phi^{001})$ and $\phi^{(2)}_i = (\phi^{200}, \phi^{020}, \phi^{002}, \phi^{011}, \phi^{101}, \phi^{110})$,
\begin{align}
\phi_i^{(1), r} = Z^{(1)}_{ij} \phi_j^{(1), \text{lat}},
&&
\phi_i^{(2), r} = Z^{(2)}_{ij} \phi_j^{(2), \text{lat}}. 
\end{align}

The renormalization factor $Z^{f_N}$ and the renormalization matrices $Z^\lambda_{ij}, Z^{(1)}_{ij}$ and $Z^{(2)}_{ij}$ have been calculated 
in \cite{Gockeler:2008we,Kaltenbrunner:2008zz}. 
There, the matching of the lattice data to a kind of RI-MOM scheme has been performed non-perturbatively, 
and the matching of the RI-MOM scheme to the $\overline{\text{MS}}$ scheme has been calculated in one-loop perturbation theory with the help of ``naive'' dimensional 
regularization that has certain shortcomings, cf.~\cite{Kraenkl:2011qb}. 
We use these results for our present study.

In \cite{Gockeler:2008we,Kaltenbrunner:2008zz}, the renormalization matrices are only given for lattices of size up to $24^3 \times 48$. 
But since there seems to be no significant volume dependence (the values for the $16^3 \times 32$ and $24^3 \times 48$ lattices agree within error bars), 
we felt comfortable to use their renormalization matrices for the $24^3 \times 48$ lattice also for our larger lattices.

\section{Data Analysis} \label{sec:data}

\subsection{Ensembles used}

The calculations in this paper have been done using the Wilson gauge action and $n_f = 2$ non-perturbatively improved Wilson (Clover) fermions. 
A list of the ensembles used is given in Table~\ref{Table:ListOfLattices}.
We would like to highlight that we have now analyzed ensembles with pion masses of $151\ \text{MeV}$, very close to the physical value.   
Hence the older ensembles used in~\cite{Gockeler:2008xv,Braun:2008ur,Braun:2009jy}, with large pion masses $m_\pi \gtrsim 450\ \text{MeV}$ 
can be neglected altogether. 
Another important improvement is that we have generated data for different lattice volumes 
(three volumes for $\beta = 5.29, \kappa = 0.13632$) and lattice spacings 
(three spacings for $m_\pi \approx 280\ \text{MeV}$) which allows us to quantify finite volume and discretization effects.
To set the scale, we use the Sommer parameter $r_0 = 0.50\ \text{fm}$~\cite{Bali:2012qs, Fritzsch:2012wq}. 

\begin{table}
\renewcommand{\arraystretch}{1.2}
\caption{Ensembles used for this work.  } \label{Table:ListOfLattices}
\begin{ruledtabular}
\begin{tabular}{lcccr}
\multicolumn{1}{c}{$\kappa$} & $m_\pi /$ MeV & Size & $m_\pi L$ & Number of\\
 & & & & configs.\footnote{The number of measurements per configuration is shown in parentheses. \\ 
$^\dagger$These ensembles were generated on the QPACE systems, financed primarily by the SFB/TR 55, while the others were generated earlier within the QCDSF collaboration. \\ 
$^*$For these ensembles, we have computed only the $\mathcal{N}_1$ interpolator and thus we do not use them for the analysis of the negative parity states. }\\
\hline \multicolumn{5}{c}{$\beta = 5.20, a = 0.0813\ \text{fm}, a^{-1} = 2427\ \text{MeV}$} \\
\hline 0.13596$^\dagger$ & 280 & $32^3 \times 64$ & 3.69 & $1999 (\times 4)$ \\
\hline \multicolumn{5}{c}{$\beta = 5.29, a = 0.0714\ \text{fm}, a^{-1} = 2764\ \text{MeV}$} \\
\hline 0.13620$^\dagger$ & 428 & $24^3 \times 48$ & 3.71 & $1991 (\times 2)$ \\
0.13620$^\dagger$ & 423 & $32^3 \times 64$ & 4.89 & $2000 (\times 2)$\\
0.13632$^*$ & 295 & $32^3 \times 64$ & 3.42 & $950 (\times 8)$ \\
0.13632 & 290 & $40^3 \times 64$ & 4.19 & $2026 (\times 2)$ \\
0.13632$^\dagger$ & 289 & $64^3 \times 64$ & 6.70 & $  1237 (\times 2)$ \\
0.13640$^*$ & 160 & $48^3 \times 64$ & 2.77 & $3499 (\times 2)$ \\
0.13640$^\dagger$ & 151 & $64^3 \times 64$ & 3.49 & $1599 (\times 3)$ \\
\hline \multicolumn{5}{c}{$\beta = 5.40, a = 0.0604\ \text{fm}, a^{-1} = 3270\ \text{MeV}$} \\
\hline 0.13647$^\dagger$ & 427 & $32^3 \times 64$ & 4.18 & $2000 (\times 2)$ \\
0.13660 & 261 & $48^3 \times 64$ & 3.82 & $2178 (\times 2)$
\end{tabular}
\end{ruledtabular}
\renewcommand{\arraystretch}{1.0}
\end{table}

\subsection{Isolating physical states} \label{subsec:isophy}

A major task in any lattice data analysis is the isolation and identification of physical states. 
To suppress excited states, we have smeared the source using Wuppertal smearing \cite{Gusken:1989qx} with APE smoothed \cite{Falcioni:1984ei} links. 
We have adjusted the number of smearing steps to optimize the plateau for the proton. 

In our previous work (\cite{Braun:2008ur,Braun:2009jy} and the data points with $m_\pi > 400\ \text{MeV}$ in \cite{Braun:2010hy,Schiel:2011av}), 
we used a different smearing (Jacobi smearing \cite{Gusken:1989ad, Best:1997qp}) with a less-optimized number of smearing steps. 
The ``jump'' seen in the coupling constants at $m_\pi \approx 400\ \text{MeV}$ in \cite{Schiel:2011av} disappeared when we re-computed them with the improved smearing. 
It was, therefore, an artifact of our analysis rather than a physics effect.

The difference to these older results is of the order of $10 \%$ for the proton and up to $50 \%$ for the negative parity states in the case of the couplings; the shape
parameters are less affected. 
The lesson is that source optimization proves to be very important for calculations of this kind, i.e., for matrix elements of local operators.

To separate the positive and negative parity states, we use the parity projectors $\gamma_{\pm}$, as described above. 

For positive parity, the state that we are interested in is the nucleon. 
It has a large overlap with the (smeared) interpolator of the form $\mathcal{N} = \mathcal{N}_1 \equiv (u C \gamma_5 d) u$, 
the ``standard'' nucleon interpolator. 
Since the mass of the nucleon is significantly lower than that of excited states, it is relatively easy to isolate.

\begin{figure}
\includegraphics[width=0.46\textwidth]{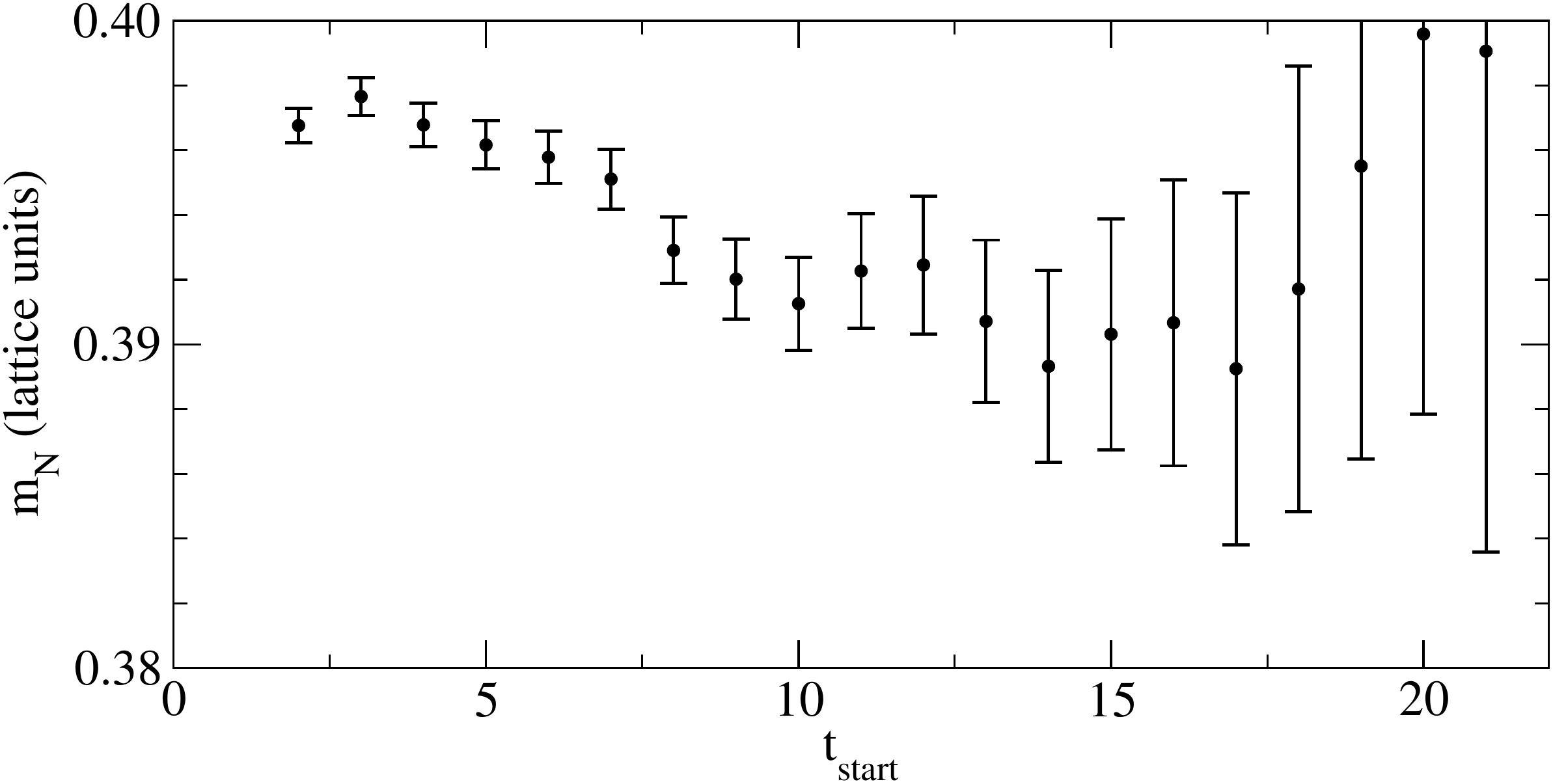}
\caption{Procedure to find a good fit range, illustrated by the example of the $\beta = 5.29, \kappa = 0.13632, 32^3 \times 64$ lattice. The end point of the fit range has been fixed to $t_\text{end} = 29$ and the starting point $t_\text{start}$ has been varied. Based on this plot, we have chosen $t_\text{start} = 9$, since variations for data points with larger starting times appear to be of statistical nature. Note the highly stretched scale, indicating the high statistical accuracy of our data.} \label{fig:plateaux}
\end{figure}

To identify a suitable time range for the fit, its start and end can be considered separately. 
The end can be determined by demanding that the influence of the backward-in-time running parity partner is negligible, 
i.e., much less than the statistical error for the state under consideration.
The starting time should be large enough that higher mass excitations are sufficiently suppressed but as small as possible to 
optimize the signal-to-noise ratio for the observables. 
In order to find optimal starting times we have generated plots for all observables like the mass plot shown in Fig.~\ref{fig:plateaux}
and made fits with fixed end point and varying starting point. We further plot the fit results with error bars
and demand that, for a good starting point, one does not observe any obvious systematic trend compared to the points with larger starting times. 
Using this starting point, the $\chi^2 / \text{d.o.f.}$ of the fit turned out to be on the order of one or smaller, indicating a good fit.

Identification of negative parity baryons on the lattice is considerably more difficult than that of the nucleon: 
In addition to the two lowest-lying $J^P = 1/2^-$ states $N^*(1535)$ and $N^*(1650)$, which only have a small mass difference, 
there are also contributions of pion-nucleon scattering states. 
 
The study~\cite{Melnitchouk:2002eg} suggested that the two negative parity states can be separated using the variational method 
with the three-quark interpolating operators $\mathcal{N}_1$ and $\mathcal{N}_2 \equiv (u C d) (\gamma_5 d)$.
In a more recent investigation using the same interpolating operators \cite{Alexandrou:2013fsu} it was found that the mass of the lower state comes out to be
very close to the sum of the nucleon and pion masses for the same lattice, suggesting it is an (S-wave) $N\pi$ scattering state. 
The higher mass state in this study has --- due to the large error bars --- a mass consistent with both $N^\ast(1535)$ and $N^\ast(1650)$
so that they could not be distinguished.
 
\begin{figure}
\setlength{\unitlength}{0.46\textwidth}
\includegraphics[width=\unitlength]{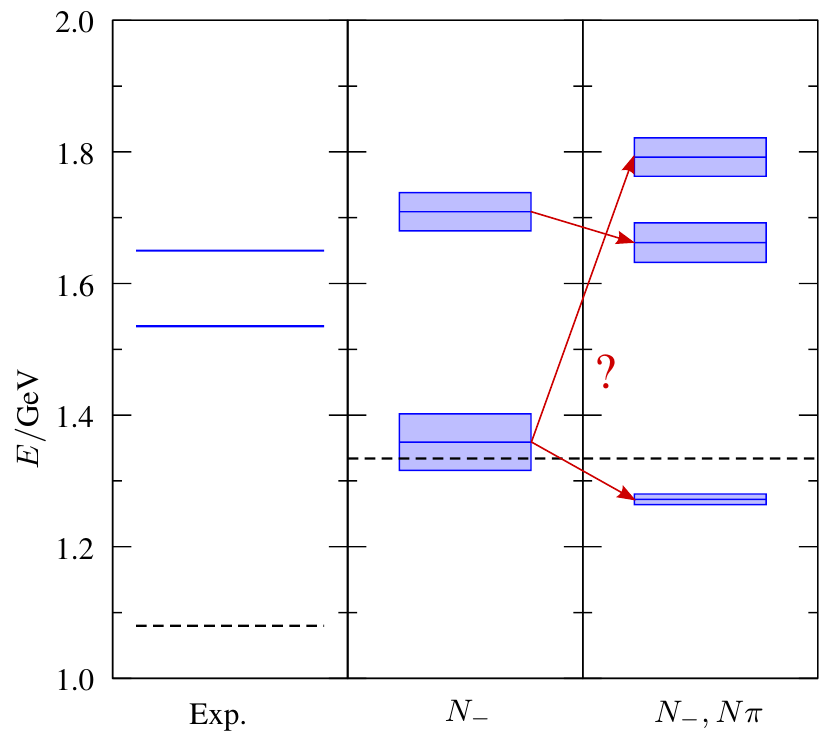}
\caption{Negative parity energy levels from experiment (left), the two-state lattice variational analysis using $\mathcal{N}_1$ and $\mathcal{N}_2$ (middle) 
and the three-state analysis including a five-quark operator (right). 
The dashed lines show the sum of the nucleon and pion masses.
This figure is taken from \cite{Lang:2012db}, with arrows added to indicate the conjectured splitting of the lower state.} \label{fig:verduci}
\end{figure}

Yet another study~\cite{Lang:2012db} uses the same interpolating operators $\mathcal{N}_1$ and $\mathcal{N}_2$ and includes in addition a third, five-quark 
interpolator to represent the nucleon-pion continuum. 
In a two-state analysis, using only the three-quark operators, their results agree with the results from \cite{Alexandrou:2013fsu}, 
yielding one state close to the nucleon-pion threshold and one heavier state. 
The full three-state analysis produces one state slightly below the nucleon pion threshold (indicating attractive interaction) 
and two heavier states that may be identified with the $N^*(1535)$ and $N^*(1650)$.
Comparing the eigenvectors of the variational basis for the two- and three-state analyses, the authors suggest that
the lower mass state of the two-state analysis splits into the $N\pi$ state and the $N^*(1650)$, while the higher mass state of the two-state analysis 
becomes the $N^* (1535)$, see Fig.~\ref{fig:verduci}.
This is also phenomenologically plausible, since the $N^*(1535)$ is not expected to mix strongly with the $N \pi$ continuum as the observed $N^*(1535)\to N \pi$ decay 
width is rather small~\cite{Beringer:1900zz}.   

Due to the high cost of five-quark interpolators we have used only the three-quark interpolators, $\mathcal{N}_1$ and $\mathcal{N}_2$, for our analysis.
Following the identification suggested in Ref.~\cite{Lang:2012db}, cf.~Fig.~\ref{fig:verduci}, we will label the lower mass state of our two-state 
variational analysis $N^* (1650 ?)$ and the higher mass state $N^* (1535 ?)$, 
where the question marks indicate that this identification is still uncertain and requires further study. 
In the case of $N^* (1650 ?)$ we expect that there is also considerable contamination by nucleon-pion scattering states.

\begin{figure}
\setlength{\unitlength}{0.46\textwidth}
\includegraphics[width=\unitlength]{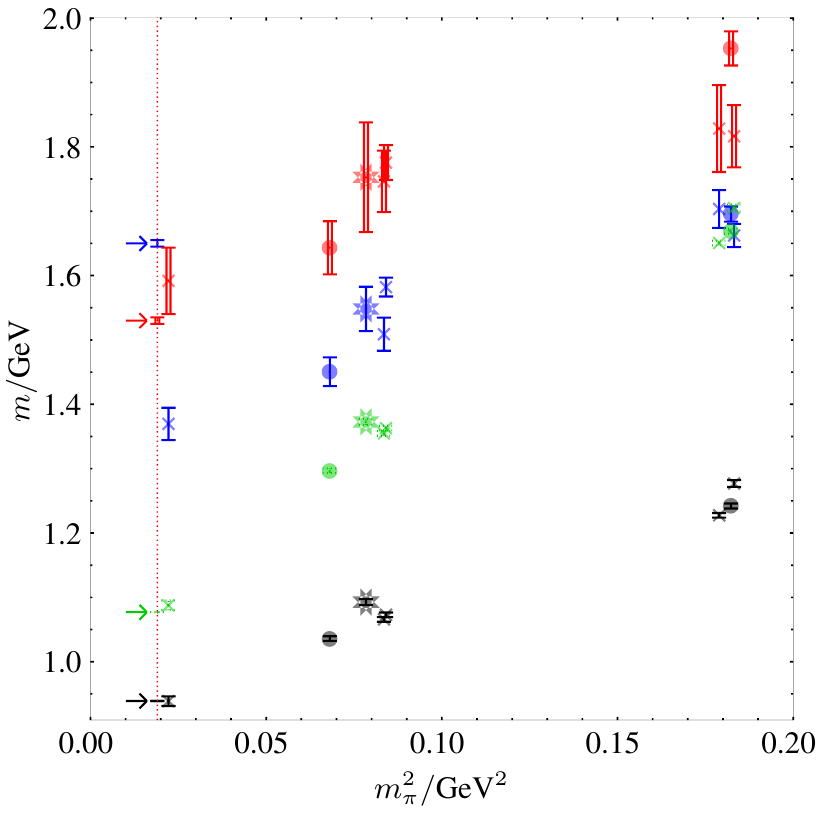}
\caption{\label{fig:masses}Masses of the nucleon (black), $N^*(1650?)$ (blue) and $N^*(1535?)$ (red, double line) as a function of the pion mass. 
The sum of the nucleon and pion masses (green, dotted error bars) is shown for comparison.
The crosses, circles and stars designate $\beta=5.29$, $\beta=5.40$ and $\beta=5.20$ data points, respectively.
The experimental values at the physical point (vertical dotted red line) are highlighted by an arrow.}
\end{figure}

The masses that we find for the nucleon and the negative parity states are shown in Fig.~\ref{fig:masses}. 
The nucleon mass has been studied in more detail in \cite{Bali:2012qs} and is -- when extrapolated to the physical point -- consistent with experiment.

The mass of the higher negative parity state (labeled $N^* (1535?)$, as explained above) 
changes rather smoothly with the pion mass and is compatible with both known resonances $N^\ast(1535)$ and $N^\ast(1650)$ within the error bars. 

For the mass of the lower state $N^* (1650?)$, \cite{Alexandrou:2013fsu} and \cite{Lang:2012db} obtain a value close to the sum of the nucleon and pion masses. 
Our ensembles with $m_\pi \simeq 420\ \text{MeV}$ confirm this behavior, but at smaller pion masses, the fitted mass appears to be significantly higher than the $N \pi$ threshold. 
Whether this is due to a smaller admixture of $N \pi$ scattering states at lower pion masses or due to larger relative momentum of the nucleon and pion within the scattering state is unclear. 
To solve this puzzle and to separate the $N^*(1650?)$ from $N \pi$ scattering states, studies with a larger variational basis, preferably with five-quark interpolators, are required, but they are too expensive at present.
Meanwhile, the identification of the negative parity states should be regarded with caution.

\subsection{Autocorrelations}

Since lattice QCD data are based on configurations which have been generated by a Markov process, 
they are subject to autocorrelations between consecutive trajectories. 
A powerful method to reduce these autocorrelations is to move the source when going from one configuration to the next: 
Using a different part of the lattice volume reduces the correlations.

To determine the remaining autocorrelations and the resulting increase in the errors, we have applied the binning method. 
For most of our observables, the binned error was only slightly, if at all, greater than the error from the ``naive'' error analysis. 
Therefore, autocorrelations were only minimal.
Merely a few observables on some ensembles showed greater autocorrelation effects and in the worst case, every other configuration was still statistically independent.  

\setlength{\unitlength}{0.43\textwidth}
\begin{figure*}
\includegraphics[width=\unitlength]{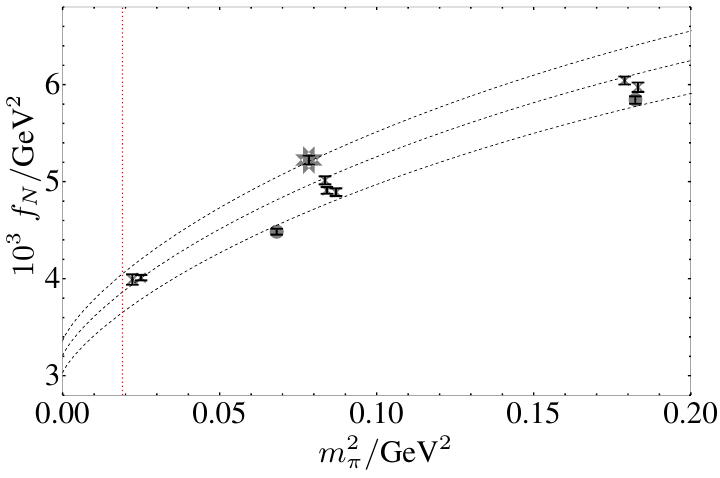}
\hspace{0.3cm}
\includegraphics[width=\unitlength]{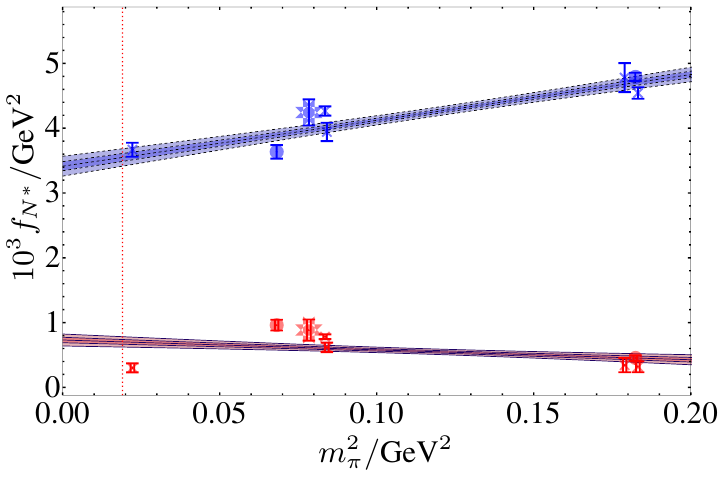}
\caption{
%
Chiral extrapolations of the wave functions at the origin $f_N$, $f_{N^*}$ for the nucleon [left panel] and the negative parity resonances $N^*(1535?)$ (red, double line) and $N^*(1650?)$ (blue) [right panel]. 
Circles correspond to the lattice data for $\beta = 5.40$, crosses to $\beta = 5.29$ and stars to $\beta = 5.20$.  
The dotted lines on the left panel show the central value of the lowest order fit scaled to the three lattice spacings 
(see Subsection \ref{subsec:context}), where the lowest line is for $\beta = 5.40$, the middle one for $\beta = 5.29$ and the highest one for $\beta = 5.20$. 
On the right panel, the $1 \sigma$ and $2 \sigma$ error bands of the fit are shown in red for $N^*(1535?)$ and in blue with dashed lines for $N^*(1650?)$.
The physical point is indicated by the vertical dotted red lines.}
\label{fig:fN-all}
\end{figure*}

\begin{figure*}
    \includegraphics[width=\unitlength]{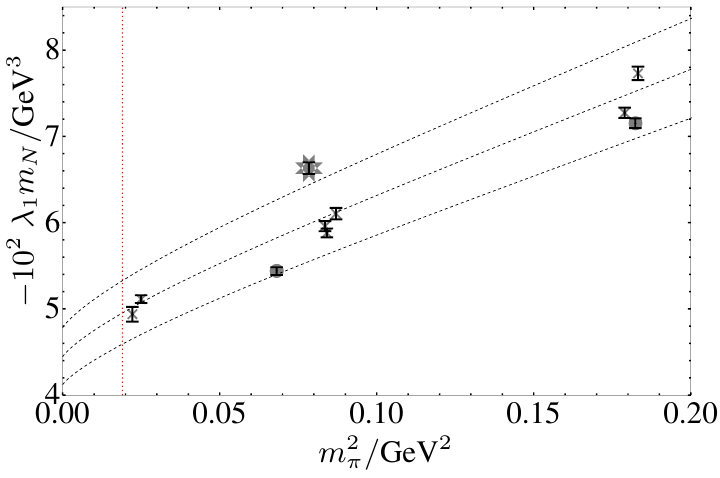}
\hspace{0.3cm}
    \includegraphics[width=\unitlength]{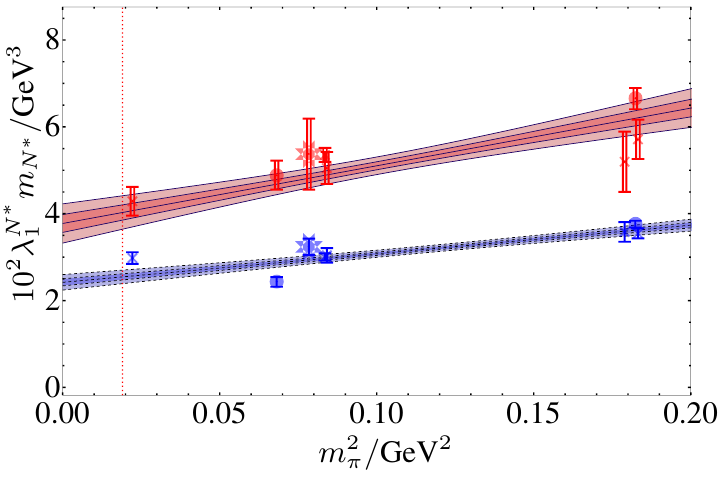}
\\[0.3cm]
    \includegraphics[width=\unitlength]{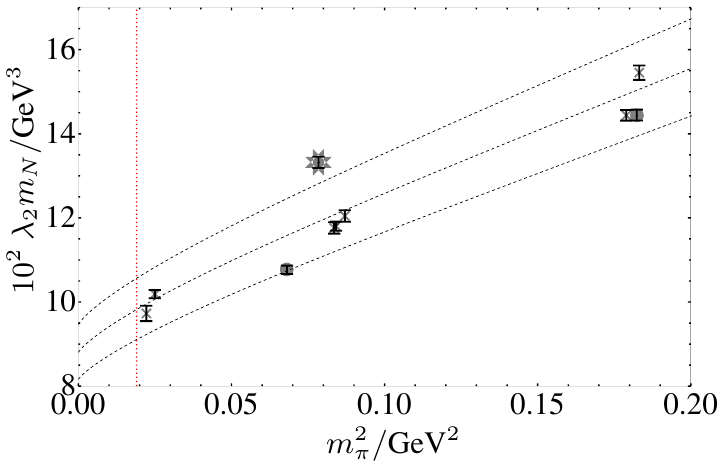}
\hspace{0.3cm}
    \includegraphics[width=\unitlength]{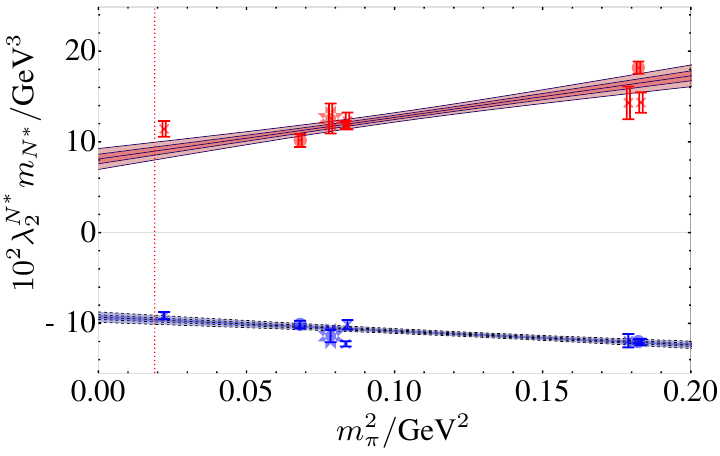}
\caption{
Chiral extrapolations of the normalization constants of the twist-4 DAs $\lambda_{1,2}$, $\lambda^{N^\ast}_{1,2}$ 
for the nucleon [left panel] and negative parity resonances [right panel]. 
The identification of the curves and the data points is the same as in Fig.~\ref{fig:fN-all}.}
\label{fig:lambdas-all}
\end{figure*}

\subsection{Chiral and infinite volume extrapolations}

The chiral extrapolations to the physical point and to infinite volume 
for the couplings $f_N$ and $\lambda_{1,2}$ are shown in Figs.~\ref{fig:fN-all} and \ref{fig:lambdas-all} and for the 
shape parameters $\varphi_{nk}$ in Appendix~\ref{App:ChiPT}.    
They have been handled differently for the nucleon and the negative parity states.

For the nucleon, extrapolation formulae for both the leading and next-to-leading twist normalization constants based on chiral perturbation theory 
($\chi$PT) are available from Ref.~\cite{Wein:2011ix}. For the  next-to-leading twist parameters we used the
combinations $m_N \lambda_1$ and $m_N \lambda_2$ in the fits, which are more natural from a $\chi$PT point of view as compared to the 
couplings themselves.

Our extrapolation formulae for the moments of the leading twist distribution amplitude are new results. 
Details of their calculation can be found in Appendix~\ref{App:ChiPT}. 
All expressions were obtained in leading one-loop covariant baryon $\chi$PT
and include correction terms for finite volume effects. 

We have fit our data with these extrapolation formulae and quote our final results for $m_\pi \rightarrow m_\pi^\text{phys}$ and $V \rightarrow \infty$. 
We have also checked the $\beta = 5.29, \kappa = 0.13632$ ensembles (where we have three different volumes) for residual finite volume effects, but have concluded that the remaining small discrepancies between the three data points must be of statistical nature.

\setlength{\unitlength}{0.4\textwidth}
\begin{figure*}
\includegraphics[width=\unitlength]{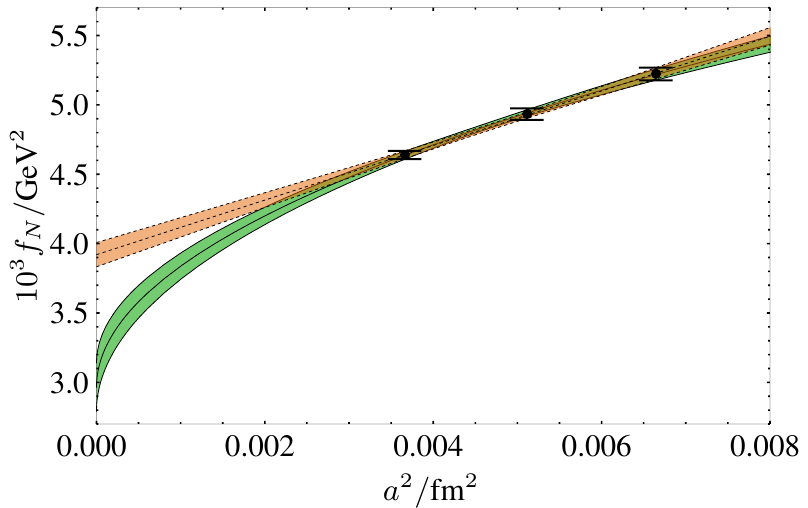} \\[0.3cm]
\includegraphics[width=\unitlength]{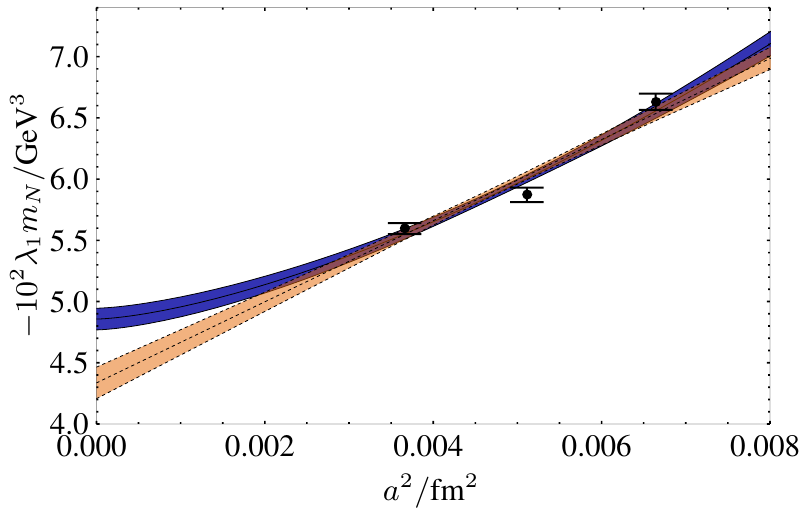}
\hspace*{0.3cm}
\includegraphics[width=\unitlength]{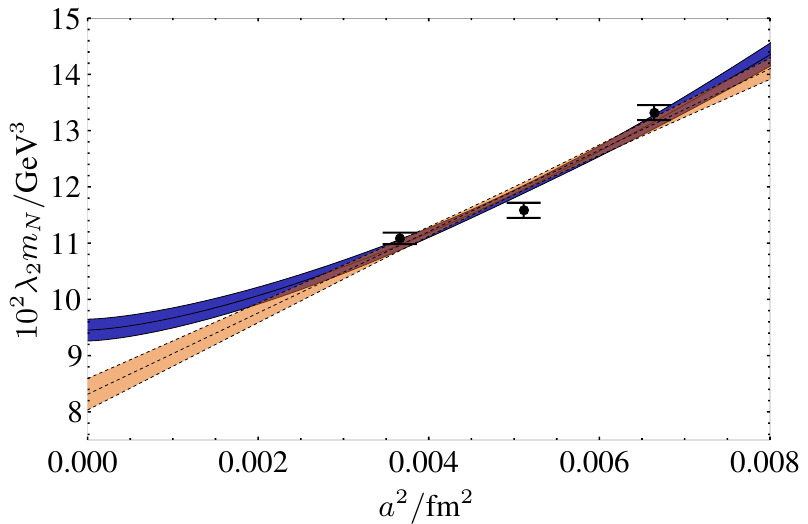}
\caption{Continuum extrapolation of the couplings $f_N$ [top], $\lambda_1$ [bottom left] and $\lambda_2$ [bottom right] using the largest volume data 
 for $m_\pi \simeq 280\ \text{MeV}$. The shaded areas correspond to 1$\sigma$ statistical error bars for the 
linear extrapolation (green, $f_N$ only), quadratic extrapolation (orange, dashed lines) and cubic extrapolation (blue, $\lambda_1$ and $\lambda_2$ only). 
} 
\label{fig:continuumextrapolation}
\end{figure*}

For the negative parity states, extrapolation formulae based on chiral perturbation theory do not exist yet. 
Therefore, we have used naive (linear) extrapolations to the physical point.
Given that our smallest pion mass is already very close to the physical value, 
the deviation of the linear extrapolation from more sophisticated approaches should be marginal.

Since we have analyzed the negative parity states for at most two volumes per $\beta$ and $\kappa$, 
a consistent study of finite volume effects for $N^*(1535?)$ and $N^*(1650?)$ is not possible. 
However, the relatively small finite volume effects that were observed for the nucleon suggest that the finite volume effects 
for the negative parity states should be reasonably small as well, i.e., at most of the order of the statistical error.

\subsection{Continuum extrapolation } \label{subsec:context}

We have analyzed ensembles with three lattice spacings, $a = 0.0813\ \text{fm}$ (corresponding to $\beta = 5.20$), 
$a = 0.0714\ \text{fm}$ ($\beta = 5.29$) and $a = 0.0604\ \text{fm}$ ($\beta = 5.40$).

\setlength{\unitlength}{0.4\textwidth}
\begin{figure*}
    \includegraphics[width=\unitlength]{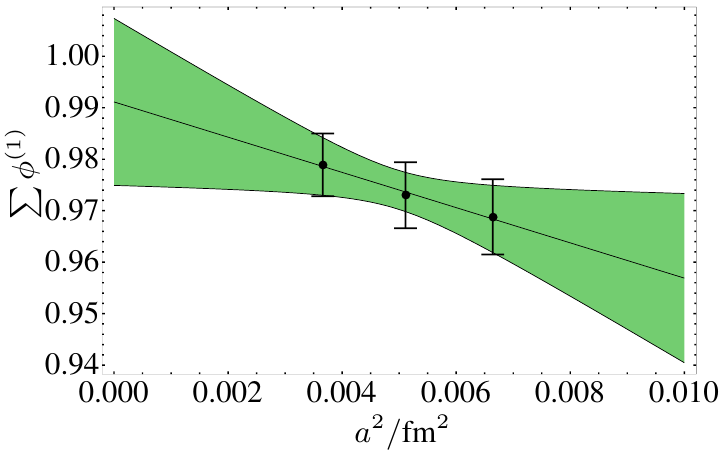}
\hspace{0.3cm}
    \includegraphics[width=\unitlength]{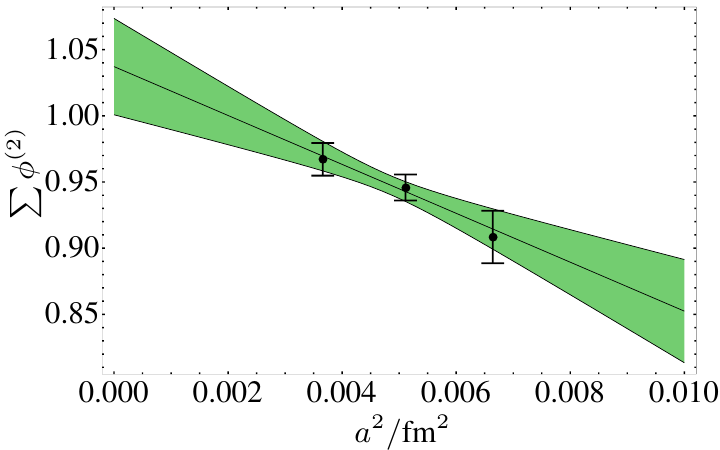}
\caption{Check of the momentum conservation constraints Eq.~(\ref{eq:sumfis}) for the nucleon as a function of lattice spacing $a$. 
For each $a$ we have used the largest volume at $m_\pi \simeq 280\ \text{MeV}$. 
} 
\label{fig:nucleonmomcons}
\end{figure*}

For $f_N, \lambda_1$ and $\lambda_2$ of the nucleon, the statistical accuracy is so high that discretization effects can be observed,
see Fig.~\ref{fig:continuumextrapolation}. 
Since the exact form of the finite $a$ corrections is unknown, we have treated the continuum extrapolation as follows: 
For $f_N$, we have tried two extrapolations, one with a linear dependence and one with a quadratic dependence on $a$, 
fitting the constants $c_N^{(1)}$ and $c_N^{(2)}$ simultaneously with the low-energy constants in
\begin{align}
f_N^{(1)} (m_\pi, a) &= f_N (m_\pi) (1 + c_N^{(1)} a), \nonumber \\
f_N^{(2)} (m_\pi, a) &= f_N (m_\pi) (1 + c_N^{(2)} a^2), \nonumber
\end{align}
where $f_N (m_\pi)$ is the $\chi$PT formula for $f_N$ and the volume dependence is suppressed for brevity. 
Both fits were almost equally good, which can be attributed to the small leverage of our three lattice spacings. 
Therefore, it is not possible to decide which fit is more accurate.
As the central value of our final result, we quote the average of $f_N^{(1)}(m_\pi^\text{phys}, 0)$ and $f_N^{(2)}(m_\pi^\text{phys}, 0)$ and as uncertainty in the continuum extrapolation one half of the difference between the two fit results.

For $\lambda_1$ and $\lambda_2$, we know that there are no $\mathcal{O}(a)$ effects, since there are no dimension $11/2$ operators in the $\tau^{\underline{4}}_1$ representation which could give rise to corrections linear in $a$, cf. Table~\ref{Table:IrredTrans}.
Therefore, we have tried extrapolations with a quadratic and a cubic dependence on $a$,  
\begin{align}
\lambda_{1,2}^{(2)} (m_\pi, a) &= \lambda_{1,2} (m_\pi) (1 + c_{1,2}^{(2)} a^2), \nonumber \\
\lambda_{1,2}^{(3)} (m_\pi, a) &= \lambda_{1,2} (m_\pi) (1 + c_{1,2}^{(3)} a^3). \nonumber 
\end{align}
Again, both fits were almost equally good and we quote the average and one half of the difference of the two fits as our central value and uncertainty of the continuum extrapolation, respectively.

Of course, also a combination of linear and quadratic corrections for $f_N$ (quadratic and cubic for $\lambda_{1,2}$) is possible 
and, with only three lattice spacings available,  will yield results with enormous uncertainties for $a \rightarrow 0$. 
Therefore, additional finer lattices will be required for a more reliable analysis of the discretization effects.

In turn, the statistical errors for the shape parameters $\varphi_{nk}$ are so large that no clear discretization effects could be observed. 
This does not imply, however, that there are no significant effects for these quantities 
and an uncertainty due to the continuum extrapolation of at least the order of the statistical error should be assumed.

An indirect argument for the consistency of the continuum extrapolation for the relevant matrix elements of the operators including derivatives
acting on the quark field can be obtained by the verification of the energy conservation relations (\ref{eq:constraints})
for the sums of first and second moments: 
\begin{align} 
\sum \phi^{(1)} &\equiv \phi^{100} + \phi^{010} + \phi^{001}, \nonumber \\
\sum \phi^{(2)} &\equiv \phi^{200} + \phi^{020} + \phi^{002} + 2 \left( \phi^{011} + \phi^{101} + \phi^{110} \right). 
\label{sumsmoments}
\end{align}
It follows from Eq.~(\ref{eq:constraints}) that these sums should be equal to one in the continuum limit,
\begin{align} \label{eq:sumfis}
\sum \phi^{(1)} = 1, \qquad \sum \phi^{(2)} = 1\,,
\end{align}
and the deviations (due to discretization errors in the Leibniz rule for derivatives) are a good measure for 
the discretization artifacts.

Since the shape parameters $\varphi_{nk}$ are extracted from differences of matrix elements corresponding to the moments $\phi^{lmn}$,
they have much larger statistical errors than the moments themselves and especially the sums of the moments in Eqs.~(\ref{sumsmoments}), which can be determined with high precision. 
These sums are plotted for the three available lattice spacings using the largest volume data 
for $m_\pi \simeq 280\ \text{MeV}$ in Fig.~~\ref{fig:nucleonmomcons}.
It is seen that the deviations are not large and the continuum extrapolated values fulfill the energy conservation constraints within the statistical 
accuracy, at the percent level for the first and 2-3\% for the second moments. These results are very encouraging and suggest that the continuum 
extrapolation is under control. 

\section{Final Results} \label{sec:res}

The final results for the normalization constants and shape parameters of the nucleon and the two lowest negative parity states, $N^*(1535?)$ and $N^*(1650?)$,
are shown in Table~\ref{table:finalres}. 
The question marks are a reminder that the identification of the results with physical negative parity resonances needs
further study and in particular we expect that the numbers for $N^*(1650?)$ include significant contributions from the pion-nucleon continuum.
For each state, the normalization constants and the moments $\phi^{lmn}$ were fit simultaneously. 
The shape parameters were then determined from the $\phi^{lmn}$ using Eqs.~(\ref{eq:cnk}).

\begin{table}
\renewcommand{\arraystretch}{1.2}
\caption{\label{table:finalres} The final results of the normalization constants and shape parameters of the nucleon and 
negative parity resonances, $N^*(1535?)$ and $N^*(1650?)$, at the scale $\mu^2=4$~GeV$^2$. 
The first error is the combined statistical error and the one due to chiral and infinite volume (for the nucleon only) extrapolation.
The second error (for the nucleon couplings) is the uncertainty of the continuum extrapolation.}
\begin{ruledtabular}
\begin{tabular}{cD{.}{.}{2.9}D{.}{.}{2.6}D{.}{.}{2.7}}
& \multicolumn{1}{c}{Nucleon} & \multicolumn{1}{c}{$N^* (1535?)$} & \multicolumn{1}{c}{$N^* (1650?)$} \\ \hline
$10^3 f_N / \text{GeV}^2$         & 2.84(1)(33)  & 0.70(4)   & 3.55(6)   \\
$10^2 \lambda_1 m / \text{GeV}^3$ & -3.88(2)(19) & 4.02(18)  & 2.54(7)   \\
$10^2 \lambda_2 m / \text{GeV}^3$ & 7.69(4)(37)  & 8.97(45)  & -9.60(23) \\
$\varphi_{10}$                    & 0.029(7)     & 0.28(12)  & 0.154(26) \\
$\varphi_{11}$                    & 0.030(4)     & -0.86(10) & 0.109(15) \\
$\varphi_{20}$                    & -0.01(8)     & 1.7(14)   & -0.07(34)\\
$\varphi_{21}$                    & -0.06(11)    & -2.0(18)  & -0.19(40)\\
$\varphi_{22}$                    & -0.02(14)    & 1.7(26)   & 0.10(63)
\end{tabular}
\end{ruledtabular}
\renewcommand{\arraystretch}{1.0}
\end{table}

The following extrapolations have been performed: chiral extrapolation to the physical pion mass (for all quantities), 
infinite volume extrapolation (only for the nucleon) and the continuum extrapolation (only for the nucleon normalization constants).
It is seen that the continuum extrapolation is the single largest source of uncertainties for the nucleon normalization constants. 
For the negative parity states, on the other hand, the results for the different lattice spacings agree within the errors.
The uncertainty in their normalization constants related to the continuum extrapolation can be expected on general grounds to be of the same order of magnitude as for the nucleon.
For the shape parameters, we expect the error due to the continuum extrapolation to be of the same order or smaller than the 
shown statistical error.

\begin{table*}
\renewcommand{\arraystretch}{1.2}
\caption{\label{table:shapecomparison} 
Comparison of our results for the nucleon shape parameters to the existing models.
The values are given at a renormalization scale $\mu^2 = 2\ \text{GeV}^2$.}
\begin{ruledtabular}
\begin{tabular}{rdddddddddd}
& \multicolumn{1}{c}{this work} & \multicolumn{1}{c}{KS} & \multicolumn{1}{c}{CZ} & \multicolumn{1}{c}{COZ} & \multicolumn{1}{c}{SB} & \multicolumn{1}{c}{BK} & \multicolumn{1}{c}{BLW} & \multicolumn{1}{c}{ABO1} & \multicolumn{1}{c}{ABO2} \\ \hline
$\varphi_{10}$ &  0.030(7) & 0.144  & 0.191  & 0.163   & 0.152 & 0.0357 & 0.0534 & 0.05       & 0.05       \\
$\varphi_{11}$ &  0.031(4) & 0.169  & 0.252  & 0.194   & 0.205 & 0.0357 & 0.0664 & 0.05       & 0.05       \\
$\varphi_{20}$ &  -0.01(9) & 0.56   & 0.32   & 0.41    & 0.65  & 0.000  & 0.000  & 0.075(15)  & 0.038(15)  \\
$\varphi_{21}$ &  -0.06(12) & -0.01  & 0.03   & 0.06   & -0.27 & 0.000  & 0.000  & -0.027(38) & -0.018(37) \\
$\varphi_{22}$ &  -0.02(15) & -0.163 & -0.003 & -0.163 & 0.020 & 0.000  & 0.000  & 0.17(15)   & -0.13(13)
\end{tabular}
\end{ruledtabular}
\renewcommand{\arraystretch}{1.0}
\end{table*}

\begin{figure}
\includegraphics[width=0.46\textwidth]{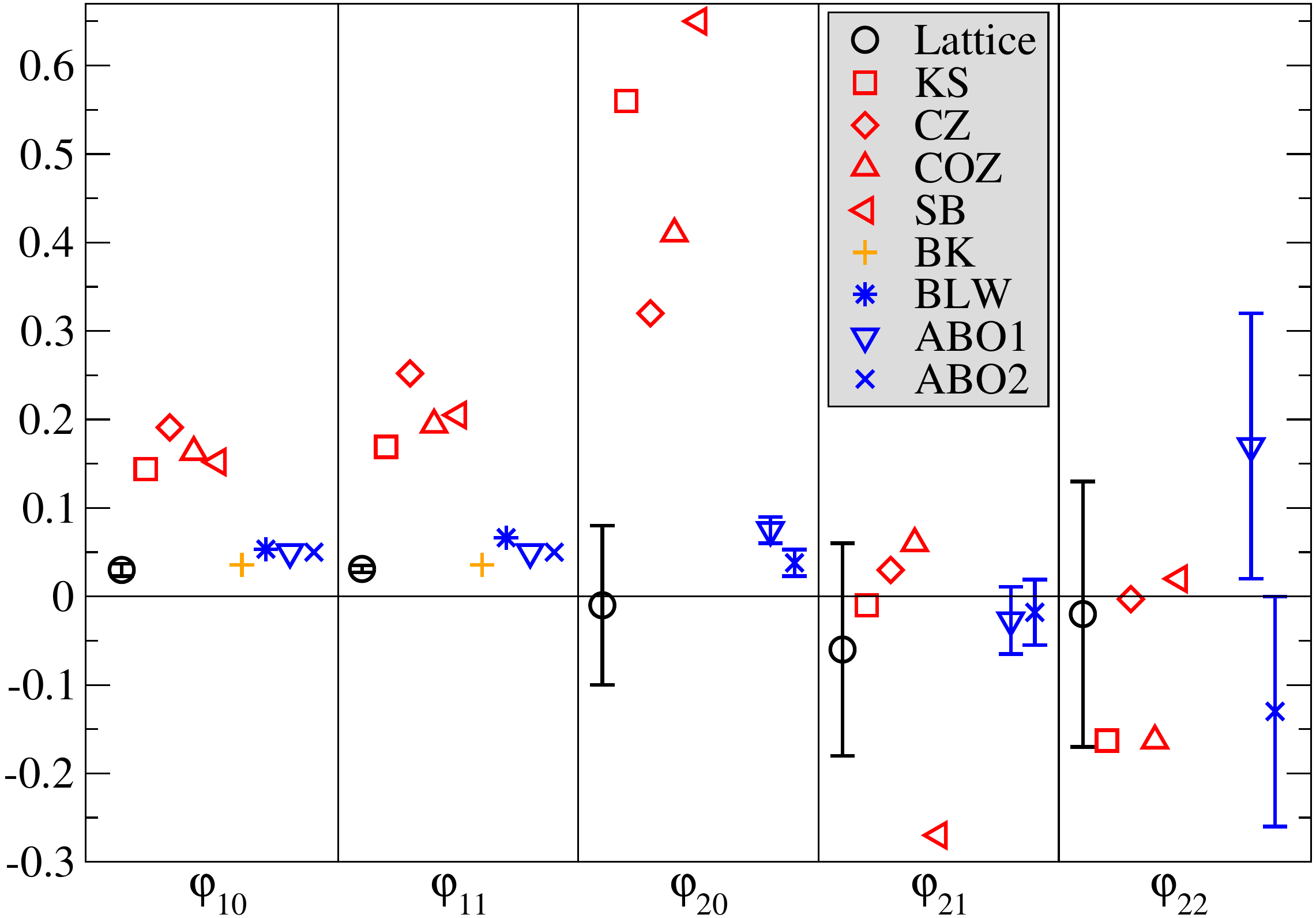}
\caption{Comparison of our results for the nucleon shape parameters (black circles) to QCD sum rule predictions (red symbols), 
light-cone sum rules (blue symbols) and the BK model (orange crosses).} 
\label{Plot:shapecomparison}
\end{figure}

Our result for $\lambda_1^N$ appears to be in a very good agreement with the next-to-leading order QCD sum rule calculation 
$m_N \lambda_1^N (\text{QCD-SR}) = -(3.4\pm 0.8)\cdot 10^{-2}\ \text{GeV}^3$~\cite{Gruber:2010bj}, but the wave function at the origin, 
$f_N$, comes out to be significantly below QCD sum rule estimates which give $f_N (\text{QCD-SR}) = (4.7\pm 0.7)\cdot 10^{-3}\ \text{GeV}^2$~\cite{Gruber:2010bj},   
where in both cases we have rescaled the QCD sum rule results from $\mu^2 = 1\ \text{GeV}^2$ to $\mu^2 = 4\ \text{GeV}^2$ using two-loop
anomalous dimensions, see Appendix~\ref{App:TwoLoop}.  
This result deals a further blow to all attempts to describe hard exclusive reactions involving nucleons at realistic energies 
in the classical perturbative QCD framework~\cite{Lepage:1980fj,Efremov:1979qk,Chernyak:1983ej}.

The main achievement of this study is the determination of the first order shape parameters of the DAs with significant precision.
These parameters are responsible for the global structure of the DAs in the momentum fraction space and, in particular, determine 
the average momentum fractions carried by the valence quarks:
\begin{align}
 \langle x_1\rangle  &= \frac13 + \varphi_{10}+ \frac13 \varphi_{11} \,,
\notag\\
 \langle x_2\rangle  &= \frac13 - \frac23 \varphi_{11}\,,
\notag\\
 \langle x_3\rangle  &= \frac13 - \varphi_{10}+ \frac13 \varphi_{11}\,.
\label{<x>}
\end{align}
The corresponding numbers are given in Eq.~(\ref{result:<x>}).

The approximate equality $\varphi_{10}\simeq \varphi_{11}$ for the nucleon and as a consequence
$ \langle x_2\rangle \simeq \langle x_3\rangle $ attracts attention.
This equality cannot be exact at all scales 
since $\varphi_{10}$ and $\varphi_{11}$ have different anomalous dimensions.
However, it is very 
interesting and suggests that the nucleon wave function (at low virtualities) is symmetric under the 
interchange of the two quarks coupled in the scalar ``diquark''. 
The diquark symmetry for the second order shape parameters would imply the constraint 
$$
 \varphi_{20} -5 \varphi_{21} + 2 \varphi_{22} =0\,.
$$
This relation cannot be checked with our data due to insufficient precision and should be addressed in future lattice calculations.   
 
A comparison of our results for the nucleon shape parameters to the existing estimates is 
shown in Table~\ref{table:shapecomparison} and Fig.~\ref{Plot:shapecomparison}. 
These are due to QCD sum rule calculations of  Chernyak and Zhitnitsky (CZ)\cite{Chernyak:1984bm}, King and Sachrajda (KS)\cite{King:1986wi}, 
Chernyak, Ogloblin and Zhitnitsky (COZ)\cite{Chernyak:1987nu}, and Stefanis and Bergmann (SB)\cite{Stefanis:1992nw}, light-cone sum rule calculations of nucleon electromagnetic form factors 
by Braun, Lenz and Wittmann (BLW)\cite{Braun:2006hz}, and Anikin, Braun and Offen (ABO1 and ABO2)\cite{Anikin:2013aka}, and the QCD-inspired 
model by Bolz and Kroll (BK)\cite{Bolz:1996sw}. For this table (and plot) we used the renormalization scale $\mu^2 = 2$~GeV$^2$. 
Our results clearly rule out the old QCD sum rule calculations of the first-order shape parameters
(alias the momentum fractions), but agree within errors with the parameters extracted from the light-cone sum rules
and the BK model. For the second-order parameters, our results rule out a large value of $\varphi_{20}$ found in~\cite{Chernyak:1984bm,King:1986wi,Chernyak:1987nu,Stefanis:1992nw}
but are otherwise consistent with zero (and with different models).

\setlength{\unitlength}{0.95\textwidth}
\begin{figure*}
    \includegraphics[width=\unitlength]{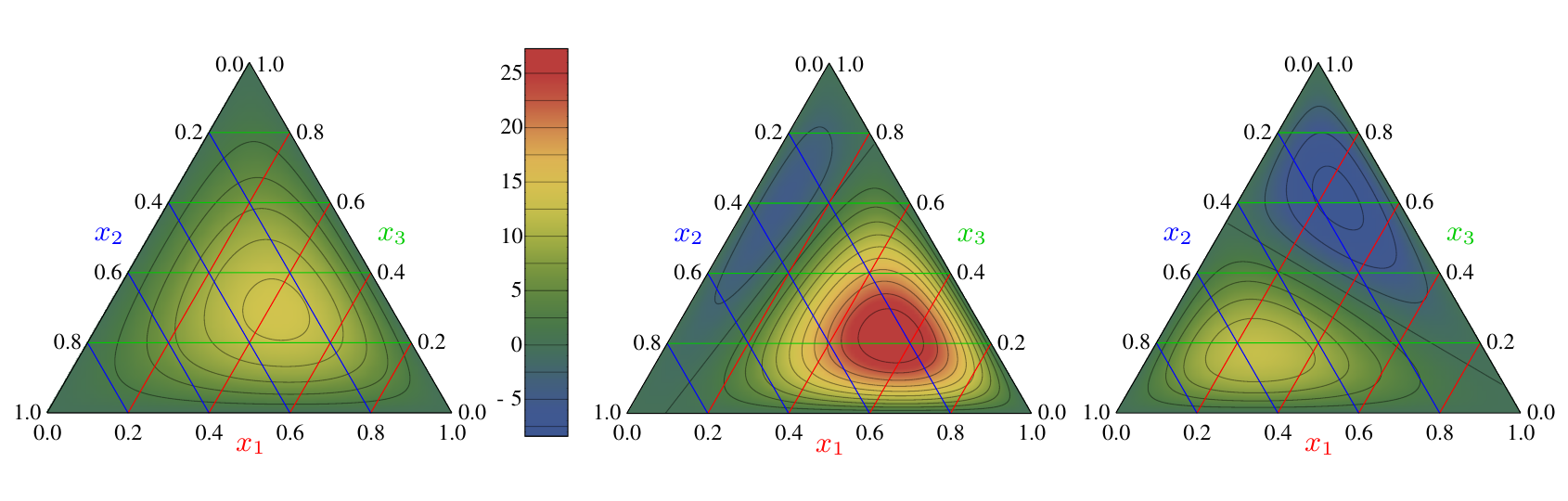}
\caption{\label{fig:barycent}Barycentric plots of the nucleon [left], $N^*(1650?)$ [center] and $N^*(1535?)$ [right] wave functions. 
Only the first moments of the distribution amplitude have been used to create these plots.}
\end{figure*}

For the negative parity states, we observe that the leading twist DA of $N^*(1650?)$ is similar to that of the nucleon, whereas 
$N^*(1535?)$ is qualitatively different: with a very small value at the origin $f_{N^\ast} \ll f_N$ and large first-order shape parameters
$\varphi_{10}^{N^\ast}$, $\varphi_{11}^{N^\ast}$ that have opposite sign to each other.
This striking difference is illustrated by the barycentric plots of the DAs in Fig.~\ref{fig:barycent}.
It can be seen that the DA of $N^*(1650?)$ (in reality, probably a mixture of $N^*(1650)$ and the pion-nucleon background) is similar to the nucleon, but 
with larger deviations from the asymptotic form. The DA of $N^*(1535?)$ appears to be completely different: 
It is approximately \emph{antisymmetric} under the exchange of the quarks in the diquark. 
This feature can be related to the observed small decay width of the $N^*(1535)$ to a pion-nucleon final state.
It is also interesting that the next-to-leading twist couplings $\lambda_{1,2}$ for the nucleon and both negative parity states are comparable, which is an indication that the quark angular momentum plays a similar role. 
The consequences of this structure for the electroproduction cross section 
of the negative parity resonances at large momentum transfer~\cite{Aznauryan:2012ba,Braun:2009jy} will be studied elsewhere.

\section{Conclusions and Outlook} \label{sec:concl}

We have presented the results of a lattice study of light-cone distribution amplitudes of the nucleon and negative parity nucleon resonances 
using  two flavors of dynamical (clover) fermions on lattices of different volumes and pion masses down to $m_\pi\simeq 150$ MeV.
Our data allow us to perform, for the first time, a reliable chiral and finite volume extrapolation of the results to the physical limit, 
and also a continuum extrapolation for some observables.
These are, to our knowledge, the first baryon structure calculations from first principles that go beyond the 
studies of the mass spectrum for the nucleon resonances.
Our results are shown in Table~\ref{table:finalres} and Fig.~\ref{fig:barycent}, and summarized in the Introduction so that we
do not need to repeat this discussion here.

The present study can be continued and improved in several directions. 
Moving to lattices with  $N_f = 2 + 1$ dynamical quarks is an obvious step.
In this way one can investigate DAs for the full baryon octet, $\Lambda, \Sigma$ and $\Xi$.    
The decay pattern of the $N^* (1535)$ (its decay fraction to $N \eta$ is $(42 \pm 10) \%$ \cite{Beringer:1900zz}) implies that the addition of 
the strange quark is important for studies of negative parity states.  
Further work is needed to improve the identification of the two lowest-lying negative parity resonances, $N^* (1535)$ and $N^* (1650)$.
The continuum extrapolation remains the largest source of errors and will be of concern as well.
There are also several other technical issues to be addressed, e.g., the matching of the RI-MOM scheme to the $\overline{\text{MS}}$ scheme 
has to be calculated to two-loop accuracy.

\begin{acknowledgments}

This work has been supported in part by the Deutsche Forschungsgemeinschaft (SFB/TR 55) and the European Union under the Grant Agreement numbers 238353 (ITN STRONGnet) and 256594 (FP7-PEOPLE-2009-RG). 
The computations were performed on the QPACE systems of the SFB/TR 55, Regensburg's Athene HPC cluster, the SuperMUC system at the LRZ/Germany and J\"ulich's JUGENE using the Chroma software system \cite{Edwards:2004sx} and the BQCD software \cite{Nakamura:2010qh}  including improved inverters \cite{Nobile:2010zz,LuscherOpenQCD}.

\end{acknowledgments}

\appendix

\section{Two-loop renormalization of the normalization constants $f_N$, 
$\lambda_{1,2}$}\label{App:TwoLoop}

For a generic nucleon coupling $f=f_N, \lambda_1, \lambda_2$
the scale dependence is given by
\begin{align}
   f(\mu) = E_f(\mu,\mu_0) f(\mu_0)
\end{align}
where
\begin{eqnarray}
&&E_{f}^{\mathrm{NLO}}(\mu,\mu _{0})=\left[ \frac{\alpha_s(\mu )}{\alpha 
_{\mathrm{s}}(\mu _{0})}\right] ^{\gamma _{f}^{(0)}/\beta_{0}}
\\
&&{}\times \left\{ 1 +\frac{\alpha_s(\mu )-\alpha_s(\mu_{0})}{2\pi \beta_0 
}
\left(\gamma_{f}^{(1)}-\frac{\beta_{1}}{2\beta_{0}}\gamma 
_{f}^{(0)}\right) \right\}.
\nonumber
\end{eqnarray}
The first two coefficients of the beta-function are
\begin{equation}
\beta _{0}=11-\frac{2}{3}n_{f}\,,\qquad
\beta_{1}=102-\frac{38}{3}n_{f}\,.
\end{equation}
Anomalous dimensions are defined such that
\begin{align}
\left[\mu^2 \frac{\partial}{\partial \mu^2} + 
\beta(\alpha_s)\frac{\partial}{\partial\alpha_s} + 
\frac12\gamma_f(\alpha_s)\right] f =0\,,
\notag\\[2mm]
\gamma_{f}(\alpha _{s}) = \gamma_{f}^{(0)}\frac{\alpha_s}{2\pi }
+\gamma _{f}^{(1)}\left( \frac{\alpha_s}{2\pi}\right)^2+\ldots.
\end{align}
The leading order anomalous dimensions are given by
\begin{align}
  \gamma_{f_N}^{(0)} = \frac{2}{3}\,,
&&
  \gamma_{\lambda_1}^{(0)} = -2\,,
&&
  \gamma_{\lambda_2}^{(0)} = -2\,.
\end{align}
The next-to-leading order (NLO) anomalous dimensions in the KM scheme~\cite{Kraenkl:2011qb} are
\begin{align}
  \gamma_{f_N}^{(1)} &= \frac{23}{9} + \frac{14}{9}\beta_0\,,
\nonumber\\
  \gamma_{\lambda_1}^{(1)} &= -\frac{19}{3} + \frac43 \beta_0\,,
\nonumber\\
  \gamma_{\lambda_2}^{(1)} &= -3 + \frac43 \beta_0\,.
\end{align}
We stress that the NLO anomalous dimensions are scheme-dependent.
Two of them, $\gamma_{\lambda_1}^{(1)}$ and $\gamma_{\lambda_2}^{(1)}$,
have been calculated also in a different scheme in Ref.~\cite{Pivovarov:1991nk}.

\section{Chiral extrapolation}\label{App:ChiPT}

We employ two-flavor baryon $\chi$PT in order to obtain a  systematic framework for the extrapolation of the nucleon distribution amplitudes 
to physical quark masses and infinite volume.
The necessary extrapolation formulae for the leading and next-to-leading twist normalization constants 
have been derived in \cite{Wein:2011ix}. For completeness we quote here the relevant expressions:
\begin{widetext}
\begin{align}
\left(\lambda_1 m_N \right)(m_{\pi}) &= \alpha^{(0)}_1 \left( 4 - \frac{m_{\pi}^2}{2(4 \pi F_{\pi})^2} \left(6 g_A^2 +(3+9 g_A^2) \ln{\frac{m_{\pi}^2}{\mu^2}} \right)  \right) + 16 \alpha^{(2)}_1(\mu) m_{\pi}^2 + \mathcal{O}(m_{\pi}^3), \nonumber \\
\left(\lambda_2 m_N \right)(m_{\pi}) &= \beta^{(0)}_1 \left( 8 - \frac{m_{\pi}^2}{(4 \pi F_{\pi})^2} \left(6 g_A^2 +(3+9 g_A^2) \ln{\frac{m_{\pi}^2}{\mu^2}} \right)  \right) + 32 \beta^{(2)}_1(\mu) m_{\pi}^2 + \mathcal{O}(m_{\pi}^3), \nonumber \\
f_N (m_{\pi}) &= \kappa^{(0)}_1 \left(1 - \frac{ m_{\pi}^2}{8(4 \pi F_{\pi})^2} \left( 6 g_A^2 +(19+9 g_A^2) \ln{\frac{m_{\pi}^2}{\mu^2}} \right)  \right) + 4 \kappa^{(2)}_1(\mu) m_{\pi}^2 + \mathcal{O}(m_{\pi}^3),  
\end{align}
\end{widetext}
where $\alpha_1^{(0,2)}, \beta_1^{(0,2)}$ and $\kappa_1^{(0,2)}$ are low-energy constants (LECs).
The dependence of the renormalized LECs $\alpha_1^{(2)}, \beta_1^{(2)}$ and $\kappa_1^{(2)}$ on the $\chi$PT-scale $\mu$ cancels the $\mu$-dependence of the logarithm $\ln (m_\pi^2 / \mu^2)$.
Finite volume corrections, which do not introduce additional low-energy constants, have also been computed.
Explicit expressions can be found in Ref.~\cite{Wein:2011ix}.

\begin{table*}
\caption{\label{LE_Chernyak_Zhitnitsky}Low-energy operators for the antisymmetric (MA) and symmetric (MS) moments of the leading twist DA grouped according to their chiral dimension $d$. We have only listed terms that contribute to the proton-to-vacuum matrix element of the operators at leading one-loop level in the limit of exact isospin symmetry and have used the shorthand $D_n \equiv n\cdot D$.}
\renewcommand{\arraystretch}{1.5}
\begin{ruledtabular}
\begin{tabular}{c c c c c c}
$d$&$k$&  $O^{\text{MA},(d)}_{k,r,LR}$ & $O^{\text{MA},(d)}_{k,r,RL}$ & $O^{\text{MS},(d),a}_{k,r,LR}$ & $O^{\text{MS},(d),a}_{k,r,RL}$ \\ \hline
$0$&$1$ & $u^{\dagger} \gamma_L \slashed{n} (i D_n)^{r+1} \Psi$ & $-u \gamma_R \slashed{n} (i D_n)^{r+1} \Psi$ & ${u^{\dagger}}^2 \tau^a u \gamma_L \slashed{n} (i D_n)^{r+1} \Psi$ & $-u^2 \tau^a u^{\dagger} \gamma_R \slashed{n} (i D_n)^{r+1} \Psi$ \\
$2$&$1$ & $\tr{\chi_+}u^{\dagger} \gamma_L \slashed{n} (i D_n)^{r+1} \Psi$ & $-\tr{\chi_+}u\gamma_R \slashed{n} (i D_n)^{r+1} \Psi$ & $\tr{\chi_+}{u^{\dagger}}^2 \tau^a u \gamma_L \slashed{n} (i D_n)^{r+1} \Psi$ & $-\tr{\chi_+}u^2 \tau^a u^{\dagger} \gamma_R \slashed{n} (i D_n)^{r+1} \Psi$ 
\end{tabular}
\end{ruledtabular}
\renewcommand{\arraystretch}{1.0}
\end{table*}

\setlength{\unitlength}{0.4\textwidth}
\begin{figure*}
\subfloat[]{
    \includegraphics[width=\unitlength]{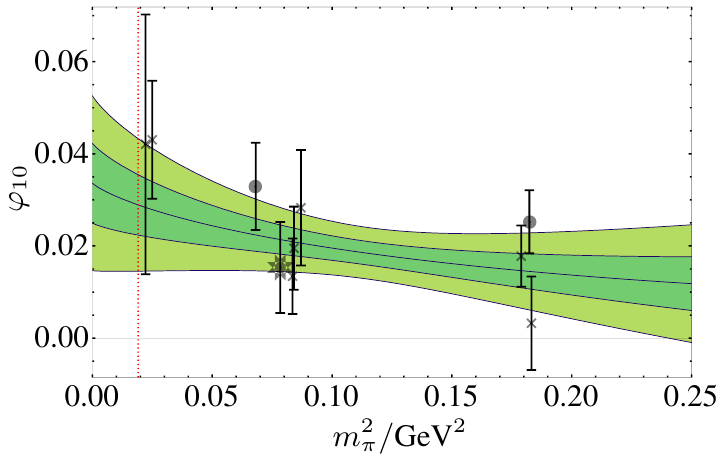}
\label{Plot:c10chiral}} 
\subfloat[]{
    \includegraphics[width=\unitlength]{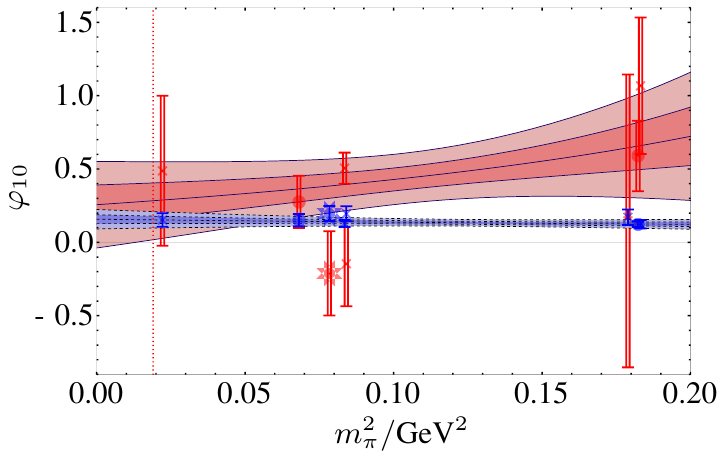}
\label{Plot:fi10star}} \\
\subfloat[]{
    \includegraphics[width=\unitlength]{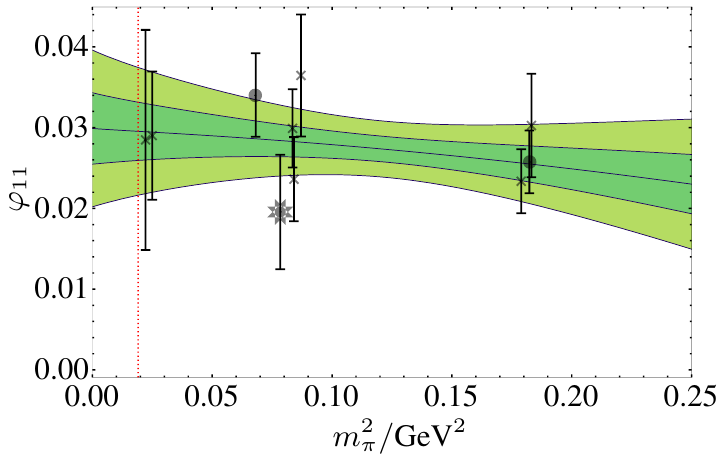}
\label{Plot:c11chiral}} \hspace{0.3cm}
\subfloat[]{
    \includegraphics[width=\unitlength]{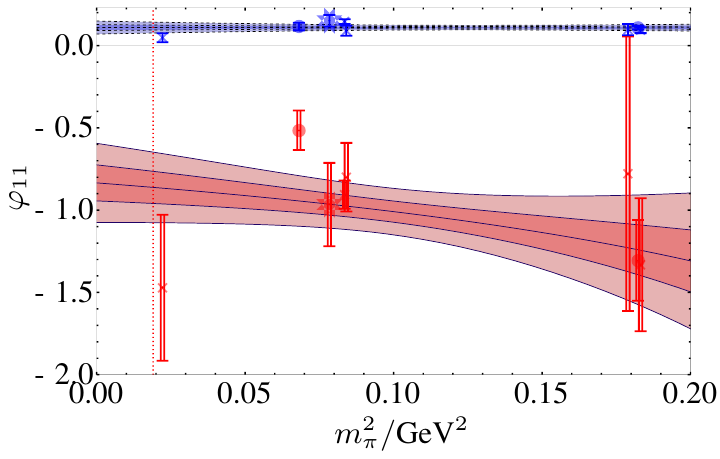}
\label{Plot:fi11star}}
\caption{
Chiral extrapolations of the shape parameters of the first order $\varphi_{10}$, $\varphi_{11}$ (\ref{expand-varphi}) 
for the nucleon [left panels] and the negative parity resonances $N^*(1535?)$ (red, double line) and $N^*(1650?)$ (blue) [right panels]. 
Circles correspond to the lattice data for $\beta = 5.40$, crosses to $\beta = 5.29$ and stars to $\beta = 5.20$.  
 The darker and lighter bands correspond to the $1 \sigma$ and $2 \sigma$ error bands of the chiral perturbation theory fit, respectively.
On the right panels, the data and error bands are shown in red for $N^*(1535?)$ and in blue with dashed lines 
(much more narrow) for $N^*(1650?)$.
The physical point is indicated by the vertical dotted red lines.}
\label{fig:phifirst}
\end{figure*}


\setlength{\unitlength}{0.40\textwidth}
\begin{figure*}
\subfloat[]{
    \includegraphics[width=\unitlength]{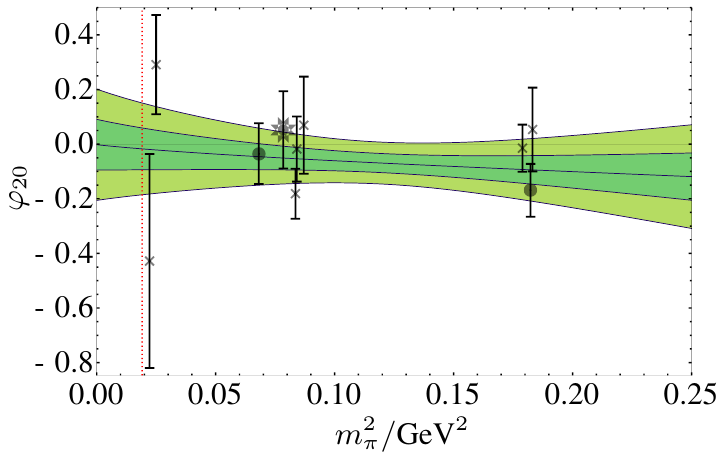}
\label{Plot:c20chiral}} \hspace{0.3cm}
\subfloat[]{
    \includegraphics[width=\unitlength]{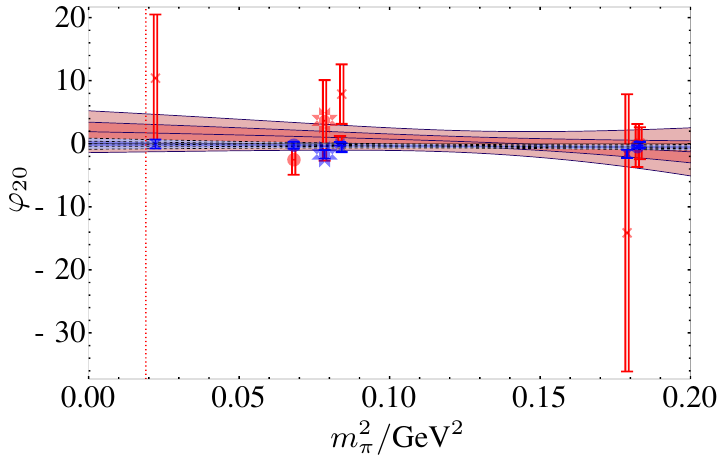}
\label{Plot:fi20star}} \\
\subfloat[]{
    \includegraphics[width=\unitlength]{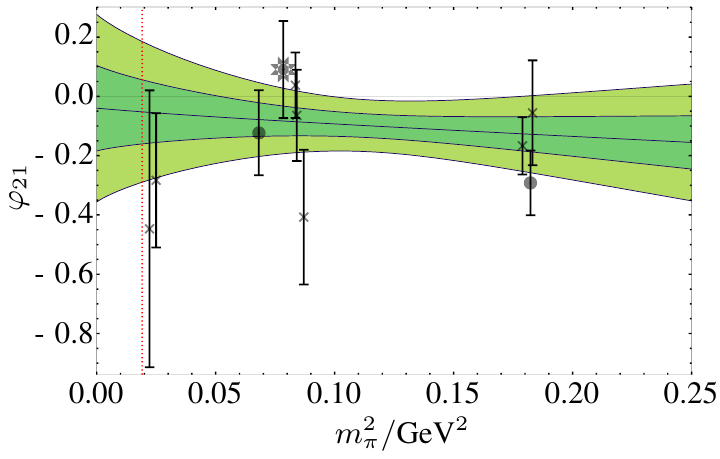}
\label{Plot:c21chiral}} \hspace{0.3cm}
\subfloat[]{
    \includegraphics[width=\unitlength]{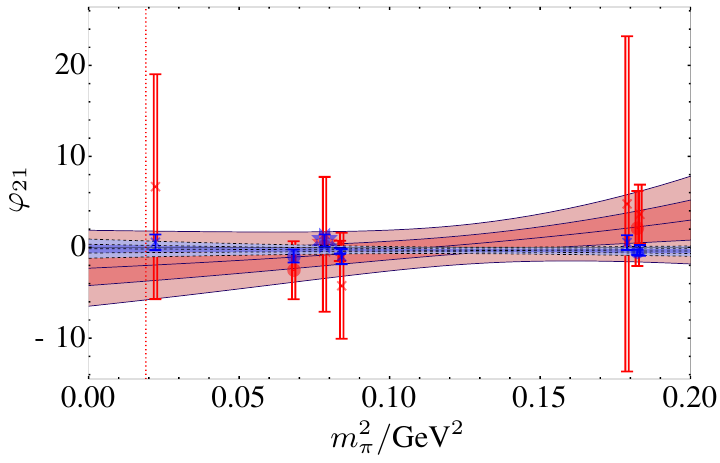}
\label{Plot:fi21star}} \\
\subfloat[]{
    \includegraphics[width=\unitlength]{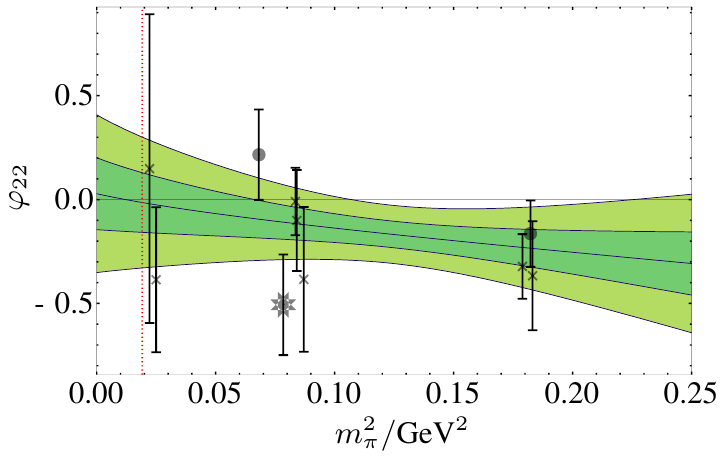}
\label{Plot:c22chiral}} \hspace{0.3cm}
\subfloat[]{
    \includegraphics[width=\unitlength]{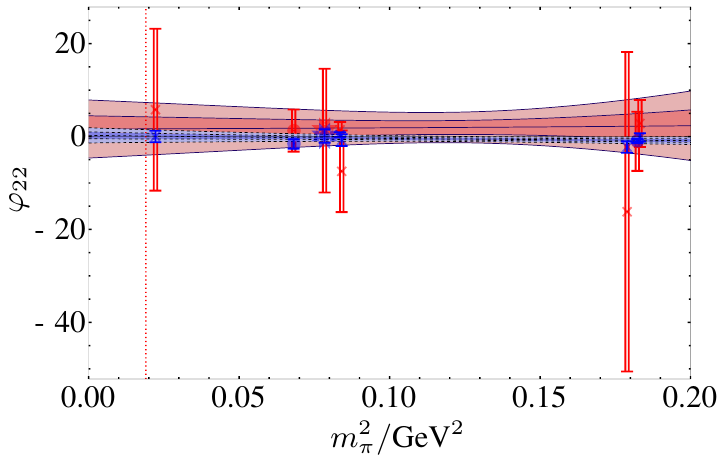}
\label{Plot:fi22star}}
\caption{
Chiral extrapolations of the shape parameters of second order  $\varphi_{20}$, $\varphi_{21}$, $\varphi_{22}$ (\ref{expand-varphi}).
The identification of the curves and the data points follows Fig.~\ref{fig:phifirst}.}
\label{fig:phisecond}
\end{figure*}

To obtain the quark mass dependence of the higher moments of the leading twist DA we follow the same procedure. 
Let us briefly describe the main steps.
To begin with, we define three-quark operators with mixed antisymmetric (MA) and mixed symmetric (MS) flavor structure 
($\text{MA} \propto uud-udu$, $\text{MS} \propto -2uud+udu+duu$)
\begin{align}
\eta^{\text{MA}}_{lmn}&=q^T_p  \left( \gamma_L \mathcal{U}^{R}_{lmn} - \gamma_R \mathcal{U}^{L}_{lmn} \right), 
\nonumber \\
\eta^{\text{MS}}_{lmn}&=\frac{4}{3}q^T_p \tau^a \left( \gamma_L \mathcal{U}^{R,a}_{lmn} - \gamma_R \mathcal{U}^{L,a}_{lmn} \right), 
\end{align}
where $q_p \equiv (1,0)^T $ projects onto the quark content of the proton and we use the notation
\begin{widetext}
\begin{align}
\mathcal{U}^{L/R}_{lmn} &\equiv \epsilon^{ijk} n_{\mu} n_{\nu} \left( \left({(i n\cdot D)^nq^i}^T_{L/R}\right)C\sigma^{\mu \rho}(i\tau^2) \left((i n\cdot D)^l q^j_{L/R}\right)\right) \sigma_{\rho}^{\ \nu}\left((i n\cdot D)^m q^k\right),
\notag \\
\mathcal{U}^{L/R,a}_{lmn} &\equiv \epsilon^{ijk} n_{\mu} n_{\nu} \left( \left({(i n \cdot D)^nq^i}^T_{L/R}\right)C\sigma^{\mu \rho}(i\tau^2)\tau^a \left((i n \cdot D)^l q^j_{L/R}\right)\right) \sigma_{\rho}^{\ \nu}\left((i n \cdot D)^m q^k\right),
\end{align}
\end{widetext}
with the quark doublet field $q \equiv (u,d)^T$. 
In the case $l=m=n=0$, the MS operator reduces to the isospin-improved Chernyak-Zhitnitsky current given in \cite{Wein:2011ix}. 
Since the transformation properties under chiral rotations are not affected by additional derivatives, 
$\gamma_L \mathcal{U}^{R,a}_{lmn}$ and $\gamma_R \mathcal{U}^{L,a}_{lmn}$ transform as $(2_L,3_R)$ and $(3_L,2_R)$, 
while $\gamma_L \mathcal{U}^{R}_{lmn}$ and $\gamma_R \mathcal{U}^{L}_{lmn}$ transform as $(2_L,1_R)$ and $(1_L,2_R)$, respectively. 
Utilizing the standard DA decomposition \cite{Braun:2000kw}, one finds that these operators project onto the moments defined in Eq.~\eqref{phi_lmn}:
\begin{align}
 \langle 0 | \eta^{\text{MA}}_{lmn} | N(p) \rangle &= f_N \frac{1}{2}\bigl(\Phi^{lmn}-\Phi^{nml}\bigr) \nonumber \\
& \quad \times (n \cdot p)^{l+m+n+1} \slashed{n} N(p), \notag \\
 \langle 0 | \eta^{\text{MS}}_{lmn} | N(p) \rangle &= f_N \frac{1}{2}\bigl(\Phi^{lmn}+\Phi^{nml}\bigr) \nonumber \\
& \quad \times (n \cdot p)^{l+m+n+1} \slashed{n} N(p).
\end{align}
The low-energy form of the operators reads
\begin{align} \label{LE_CZ_DEF}
\eta^{\text{MA}}_{lmn}&=q_p^T  \sum_{d=0}^{\infty} \sum_{k=1}^{i_d} \kappa_{k \ lmn}^{\text{MA},(d)} \left( O^{\text{MA},(d)}_{k,(l+m+n),LR } \right. \nonumber \\
&\quad \left. - O^{\text{MA},(d)}_{k,(l+m+n),RL} \right), \notag \\
\eta^{\text{MS}}_{lmn}&=\frac{4}{3}q_p^T \tau^a \sum_{d=0}^{\infty} \sum_{k=1}^{i_d} \kappa_{k \ lmn}^{\text{MS},(d)} \left( O^{\text{MS},(d),a}_{k,(l+m+n),LR } \right. \nonumber \\
&\quad \left. - O^{\text{MS},(d),a}_{k,(l+m+n),RL} \right),
\end{align}
i.e., all operators of the same symmetry class containing the same number of derivatives only differ in the LECs, since the operators built of chiral fields cannot be sensitive to the actual position of the derivatives. 
By construction, the occuring LECs obey the following constraints:
\begin{align}
\kappa_{k \ lmn}^{\text{MS},(d)} = \kappa_{k \ nml}^{\text{MS},(d)} , &&
\kappa_{k \ lmn}^{\text{MA},(d)} = -\kappa_{k \ nml}^{\text{MA},(d)} .
\end{align}
The LECs are further constrained by Eq.~\eqref{eq:constraints} 
which ensures energy-momentum conservation in plus direction 
(this reduces the number of parameters for a simultaneous fit of the 0th, 1st and 2nd moments from $20$ to $12$). 
The terms contributing to the leading one-loop calculation of the respective matrix elements in the 
limit of exact isospin symmetry can be taken from Table~\ref{LE_Chernyak_Zhitnitsky}.

One finds that the additional Lorentz indices can only come from derivatives acting on the nucleon field 
(all other possibilities are either of higher order, contain too many pion fields or are zero after contraction with the light-cone vector $n$). 
Calculating the relevant Feynman diagrams and expanding the result to the valid order in $m_\pi$, one finds:
\begin{widetext}
\begin{align}
\left(f_N \frac{1}{2} \bigl(\Phi^{lmn}-\Phi^{nml} \bigr)\right)(m_\pi) &= \kappa^{\text{MA},(0)}_{1 \ lmn} \left(1 - \frac{ m_{\pi}^2}{8(4 \pi F_{\pi})^2} \left( 6 g_A^2 +(3+9 g_A^2) \ln{\frac{m_{\pi}^2}{\mu^2}} \right)  \right) + 4 \kappa^{\text{MA},(2)}_{1 \ lmn}(\mu) m_{\pi}^2 + \mathcal{O}(m_{\pi}^3), 
\notag\\
\left(f_N \frac{1}{2} \bigl(\Phi^{lmn}+\Phi^{nml} \bigr)\right)(m_\pi) &= \kappa^{\text{MS},(0)}_{1 \ lmn} \left(1 - \frac{ m_{\pi}^2}{8(4 \pi F_{\pi})^2} \left( 6 g_A^2 +(19+9 g_A^2) \ln{\frac{m_{\pi}^2}{\mu^2}} \right)  \right) + 4 \kappa^{\text{MS},(2)}_{1 \ lmn}(\mu) m_{\pi}^2 + \mathcal{O}(m_{\pi}^3).
\end{align}
Finite volume corrections can be calculated as described in Ref.~\cite{Wein:2011ix}.

The chiral extrapolations of lattice data to the physical point and to infinite volume 
for the couplings $f_N$ and $\lambda_{1,2}$ are shown in Figs.~\ref{fig:fN-all} and \ref{fig:lambdas-all} in the text and for the 
shape parameters $\varphi_{nk}$ in Figs.~\ref{fig:phifirst} and \ref{fig:phisecond} in this Appendix. 
\end{widetext}
\bibliography{bdas_003}

\end{document}